\documentclass{aastex}
\usepackage{url}
\usepackage{spr-astr-addons}
\usepackage{graphicx}
\usepackage{amsmath}
\usepackage{rotating}
\usepackage{amssymb}
\usepackage{epsfig}

\begin{document}

\title{Dispersion in Neptune's Zonal Wind Velocities from NIR Keck AO Observations in July 2009}

\shorttitle{Neptune's Zonal Winds in 2009}
\shortauthors{Fitzpatrick, de Pater, Luszcz-Cook, Wong, \& Hammel}

\author{Patrick J. Fitzpatrick\altaffilmark{1}, Imke de
  Pater\altaffilmark{1,2}, Statia
  Luszcz-Cook\altaffilmark{1,3}, Michael
  H. Wong\altaffilmark{1}, Heidi B. Hammel\altaffilmark{4}}
\affil{Email: fitzppat@berkeley.edu (PJF)}
\altaffiltext{1}{Department of Astronomy, University of California, Berkeley, CA 94720, USA}
\altaffiltext{2}{Faculty of Aerospace Engineering, Delft University of
  Technology, 2629 HS Delft, and SRON Netherlands Institute for Space
  Research, 3584 CA Utrecht, The Netherlands}
\altaffiltext{3}{Astrophysics Department, American Museum of Natural
  History, Central Park West at 79th Street, New York, NY  10024, USA}
\altaffiltext{4}{AURA, 1212 New York Ave. NW, Suite 450, Washington, DC
  20005, USA}

  
\begin{abstract}

We report observations of Neptune made in H-(1.4-1.8 $\mu$m) and K'-(2.0-2.4 $\mu$m) bands on 14 and 16 July 2009 from the 10-m W.M. Keck II Telescope using the near-infrared camera NIRC2 coupled to the Adaptive Optics (AO) system.  We track the positions of 54 bright atmospheric features over a few hours to derive their zonal and latitudinal velocities, and perform radiative transfer modeling to measure the cloud-top pressures of 50 features seen simultaneously in both bands.

We observe one South Polar Feature (SPF) on 14 July and three SPFs on 16 July at
$\sim$65$^{\circ}$S.  The SPFs
observed on both nights are different features, consistent with the high variability of Neptune's storms.

There is significant dispersion in Neptune's zonal wind velocities about the smooth Voyager wind profile fit of Sromovsky et al., \icarus\ \textbf{105}, 140 (1993), much greater than the upper limit we expect from vertical wind shear, with the largest dispersion seen at equatorial and southern mid-latitudes.  Comparison of feature pressures vs. residuals in zonal velocity from the smooth Voyager wind profile also directly reveals the dominance of mechanisms over vertical wind shear in causing dispersion in the zonal winds.

Vertical wind shear is not the primary cause of the difference in dispersion and deviation in zonal velocities between features tracked in H-band on 14 July and those tracked in K'-band on 16 July.  Dispersion in the zonal velocities of features tracked over these short time periods is dominated by one or more mechanisms, other than vertical wind shear, that can cause changes in the dispersion and deviation in the zonal velocities on timescales of hours to days.

\end{abstract}

  
\section{Introduction}
The zonal wind velocities of the giant planets can be
derived by tracking the motions of cloud features in their
atmospheres.  Accurate tracking of the motions of cloud features in
Neptune's atmosphere was first achieved with data from the Voyager 2
spacecraft in 1989 \citep{stone98}.  \citet{sromovsky93} made a smooth fit to the zonal velocities vs. latitudes of discrete cloud features in Neptune's atmosphere which were tracked by \citet{limaye91} in Voyager 2 images.  Neptune's canonical zonal wind profile is this smooth Voyager wind profile.  The wind velocities
derived from individual features in Neptune's atmosphere since Voyager 2
are observed to remain consistent with this smooth Voyager wind
profile, with the exception of features which display significant deviation in zonal velocity, presumably due to mechanisms such as vertical wind shear, wave phenomena, or eddy motions, and which have sometimes been associated with structures such as dark spots (e.g. \citealt{hammel97, sromovsky01b,
  sromovsky01c, sromovsky02}).  \citet{sromovsky01c} found deviations from the smooth Voyager wind profile in 1998 HST observations which were consistent with observations in 1995 and 1996 (\citealt{sromovsky01b}), but they concluded that more measurements are needed to confirm a change in Neptune's zonal wind structure.

Striking dispersion and variation in zonal
wind velocities has been observed on Neptune since the Voyager era \citep{smith89}.  \citet{limaye91} found significant dispersion in zonal wind velocities, with the greatest dispersion found for more short-lived features, and at equatorial and mid-latitudes.
\citet{sromovsky93} also noted significant deviation in the zonal velocities of cloud features
about their smooth Voyager profile fit to the data of Limaye \& Sromovsky (1991). \citet{hammel97} studied the zonal motions of cloud features in 1995 HST images, along with those for features found in Voyager images (\citealt{limaye91, hammel89s}), Voyager radio occultation data (\citealt{lindal90}), 1991 HST data (\citealt{sromovsky95}), and ground-based data (see \citet{sromovsky93} Table VII) and noted that all measurements show dispersion in velocities in narrow bands of latitude, which the authors indicated as evidence for shear, wave phenomena, or other local disturbances.
\citet{sromovsky01b} found significant deviation in zonal velocities from the smooth Voyager wind profile for features tracked in HST data from 1994, 1995, and 1996. The authors found deviation in 1996 HST data to be mostly associated with a Great Dark Spot at 32$^{\circ}$N, thought to be the same dark spot seen in 1994 HST images of Neptune at 30$^{\circ}$N \citep{hammel95}. The authors associated dispersion in this region with waves propagating from this Great Dark Spot or associated standing waves.

The more recent analysis of \citet{martin12} reveals
significant dispersion in zonal wind velocities from the smooth Voyager
wind profile.  \citet{martin12} observed Neptune in
H-band with the Keck AO system for 4 hours on UT 20 and 21 August, and for $\sim$1 hour on UT 1 September, 2001.  The authors reliably measured the relative velocities of almost 200 clouds in
Neptune's atmosphere, characterizing the dispersion in
Neptune's zonal wind velocities about the smooth Voyager wind profile.  The authors found significant dispersion in zonal wind velocity (with variations as high as $\sim$500
m/s), greater than the upper limit they placed on the contribution to dispersion caused by vertical wind shear, and which they attributed primarily to
outlying transient clouds that do not move with their local mass flow.

We conduct a similar study as \citet{martin12}, except we track the
motions of Neptune's cloud features in both H- ($\sim$1.6$\mu$m) and K'-
($\sim$2.2$\mu$m) bands (in observations separated by $\sim$3 Neptune rotations), which
are sensitive to a different range of altitudes in Neptune's atmosphere,
to search for potential differences between the two wavelengths.  We also perform radiative transfer modeling to estimate the pressures of features.  Comparison of the zonal velocities of features with their estimated pressures provides a direct probe of the relative contribution of vertical wind shear to dispersion in the zonal wind velocities.  Because our observations in H-band probe a greater overall
magnitude and range of depths in Neptune's atmosphere than those in
K'-band, comparing the dispersion observed between the two filters can also
give us insight into the relative contribution of vertical wind shear to dispersion in the
zonal winds.

In Section 2 we describe our observations and data,
including our method of alignment and cylindrical projection of images.
In Section 3 we describe our method of tracking the longitude-latitude
positions of atmospheric features in our images.  Section 4
describes our results for the dispersion in Neptune's zonal winds, along
with our results for the depths of features from radiative transfer
modeling, and our observations of the South Polar Features.  In
Section 5 we discuss our results in the context of a few relevant mechanisms
which can cause dispersion in Neptune's zonal winds and we compare the
dispersion observed in H- and K'-bands on 14 and 16 July, respectively.  Finally, Section 6 summarizes
our conclusions.

\section{Data}

\subsection{Observations and Data Reduction}
We observed Neptune from the 10-m W.M. Keck II Telescope
on Mauna Kea, Hawaii, on 14 and 16 July 2009 (UT) as part of a project to study the planet's atmosphere and rings at near-infrared (NIR) wavelengths.  H-(1.4-1.8$\mu$m) and K'-(2.0-2.4$\mu$m) band images were taken using the
 narrow camera of the NIRC2 instrument, coupled to the AO system.  The 1024$\times$1024 array has a pixel scale of
 9.963$\pm$0.011 mas/pixel in this mode \citep{pravdo06}, which at
 the time of observations corresponded to a physical scale of $\sim$210 km/pixel
 at disk center.

Although we observed Neptune in both
 bands on each day, images were more frequently taken in H-band on 14
 July and in K'-band on 16 July.  On 14 July we took a total
 of 75 H-band images spanning $\sim$2.5 hours (11:20:21 - 13:47:44
 (UT)) and on 16 July we took a total of 105 K'-band images spanning
 $\sim$3.5 hours (10:54:57 - 14:20:09 (UT)).  The integration times of both
 H- and K'-band images are 60 sec, which provide enough signal to noise without saturating our
 detector, and assure that smearing due to Neptune's rotation is less than one image pixel.  Short integration times enable a dense
 sampling of images over time, allowing us to accurately identify the
 same features in successive images.  Our data were taken in $\sim$5-min
 sequences of five frames each.  Largest separations between sequences
 of images were $\sim$30 min, during which observations of photometric
 standards were carried out.

We reduced images using standard infrared data
 reduction techniques of sky subtraction, flat fielding, and
 median-value masking to remove bad pixels.  All images were corrected
 for the geometric distortion of the array using the `dewarp' routines
 provided by P. Brian
 Cameron,\footnote[1]{http://www2.keck.hawaii.edu/inst/post\_observing/dewarp/\\nirc2dewarp.pro}
 who estimates residual errors at $\lesssim$0.1 pixels.  We measured
 angular resolution with the full width at half maxima (FWHM)
 of Neptune's moons visible in our images.  We measure FWHM in H-band on
 14 July of 0.050$\pm$0.004'' and in K'-band on 16 July of
 0.049$\pm$0.005'', which are consistent with the diffraction limit of
 our telescope at 2$\mu$m \citep{vandam04}, and correspond to effective
 resolutions of $\sim$1,060 km and $\sim$1,037 km at the center of the
 disk, respectively.

Images were photometrically calibrated using the star HD201941, and were converted from units of observed flux density to units of I/F, which is defined as \citep{hammel89a}:
\begin{equation}
\frac{I}{F} = \frac{r^{2}}{\Omega}\frac{F_{N}}{F_{\odot}},
\end{equation}
where $r$ is Neptune's heliocentric distance, $\pi F_{\odot}$ is the Sun's flux density at Earth's orbit \citep{colina96}, $F_{N}$ is Neptune's observed flux density, and $\Omega$ is the solid angle subtended by a pixel on the detector.  By this definition, I/F=1 for uniformly diffuse scattering from a Lambert surface when viewed at normal incidence.

\subsection{Locations of Cloud Features}
\begin{figure*}[!h]
\includegraphics[width=\linewidth]{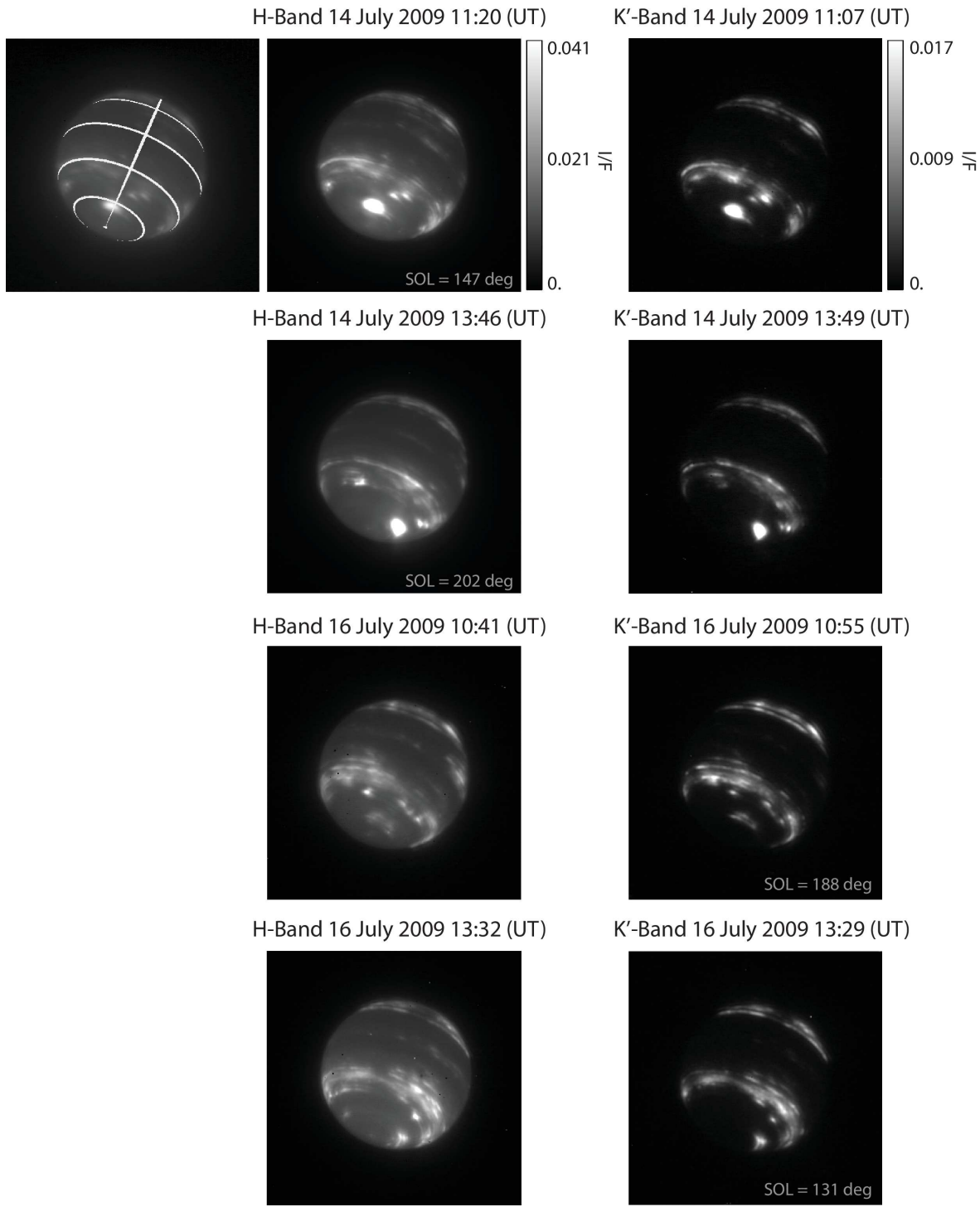}
\caption{Neptune H- (left column) and K'-band (right column)
    images taken on 14 July 2009 (top two rows) and 16 July 2009
    (bottom two rows) towards the beginning (first and third rows) and end (second and fourth
rows) of each night.  Images are shown in units of I/F according to the
scale shown in the colorbars to the right of the topmost images in each
filter.  For comparison, to the left of the 14 July H-band image taken
at 11:20 (UT) (top left panel) we show the same image with overlayed
lines of constant latitude (lines are shown at 60$^{\circ}$S,
30$^{\circ}$S, the equator, and 30$^{\circ}$N) and longitude (a line is
shown at the sub-observer longitude).  The sub-observer longitudes of
navigated images are indicated in the bottom right corners of panels.}
\label{discs}
\end{figure*}

Figure \ref{discs} shows Neptune images on
both days of observation in
H-band (left panels) and K'-band (right panels), taken towards the
beginning (1st and 3rd rows) and end (2nd and 4th rows) of each night.  On both 14 and 16 July we observed Neptune at
roughly the same longitudes (these images are separated by $\sim$3
Neptune rotations).  We
immediately note that atmospheric features tend to be distributed
preferentially along bands of constant latitude, with increased
prevalence at mid-latitudes, in agreement with previous
observations (e.g. \citealt{sromovsky01b, max03, irwin11, martin12}).  In addition to clouds at roughly the same latitudes as \citet{martin12}, we observe clouds at $\sim$40$^{\circ}$N.  In all images there is an absence of cloud features just south of the equator, in
agreement with previous observations (e.g. \citealt{limaye91,
  martin12}).  Atmospheric features vary drastically between 14 and 16 July.  Feature morphology changes so dramatically that we cannot with certainty identify the same features present on both nights, as has been noted in previous observations (e.g. \citealt{sromovsky01b, karkoschka11b}).

In all 14 July images we observe a large
bright feature centered at $\sim$65$^{\circ}$S.  In 16 July images, centered at
the same latitude, we observe three distinct features instead of
the large bright feature seen on 14 July.  Due to foreshortening
effects, these three features appear to begin to coalesce into one as
they approach the limb at the end of the night.  Features in
this latitude region have been identified in previous observations as
South Polar Features (SPFs; e.g. \citealt{hammel89b, smith89, limaye91, sromovsky93, crisp94}; Hammel \& Lockwood 1997; \citealt{sromovsky01b, sromovsky01c, rages02}; Karkoschka 2011a; \citealt{martin12}).  We
will discuss these SPFs in Section \ref{SPFs}.

In all H-band images we observe a small
bright feature at the south pole of the planet.  In K'-band we cannot see this
feature.  This feature has been observed since the Voyager era
(e.g. \citealt{smith89, limaye91, hammel07, statia10, karkoschka11a}).  As seen by \citet{limaye91}, this feature persisted in
the Voyager observations over many Neptune rotations, and the authors were tempted to suggest that it marked the planet's true rotation pole, as doing so would remove a mean meridional velocity bias which puzzled them.  \citet{martin12} assumed this `south pole dot' to mark Neptune's
true rotation pole in verifying image navigation, and found the planet center deduced from the south pole dot to agree with that determined from limb fitting within 1 image pixel (16.7$\pm$0.2 mas).  There is no a priori reason to assume the feature at the south pole to mark Neptune's true rotation pole.  In fact in previous observations a pair of south polar spots have been seen $\sim$1-2$^{\circ}$ away from the south pole (\citealt{statia10, karkoschka11a}), and the former authors suggested that these clouds may form in a region of strong convection surrounding a south polar vortex.

\subsection{Image Navigation and Cylindrical Projection}
To navigate and align our images we must determine the location of the physical center of Neptune in each image to within sub-pixel accuracy.  We do so using a multivariate nonlinear $\chi^{2}$ minimization routine
which simultaneously fits for the positions of three moons onto their
respective orbits for each disk image.  The orbit of each moon was
derived using the ephemeris generators in the Rings Node of NASA's
Planetary Data System (http://pds-rings.seti.org/).  We derive moon
orbits rather than use their individual locations given by the ephemeris
generators due to uncertainties in the latter (e.g. \citealt{jacobsen04, imke05}).  This is based on the fitting routine used to align Neptune images by \citet{statia10}.  In 14 July H-band images we fit for the simultaneous positions of
Galatea, Larissa, and Despina, and in 16 July K'-band images we fit for
the simultaneous positions of Galatea, Larissa, and Proteus.  Our mean $\left( x, y \right)$ uncertainties in the derived Neptune centers in our images (calculated from the variance modified by a factor of the reduced-$\chi^{2}$) for H- and K'-bands are (0.08 pix,0.07 pix) and (0.09 pix,0.08 pix), respectively.

\begin{figure*}[t]
\includegraphics[width=\linewidth]{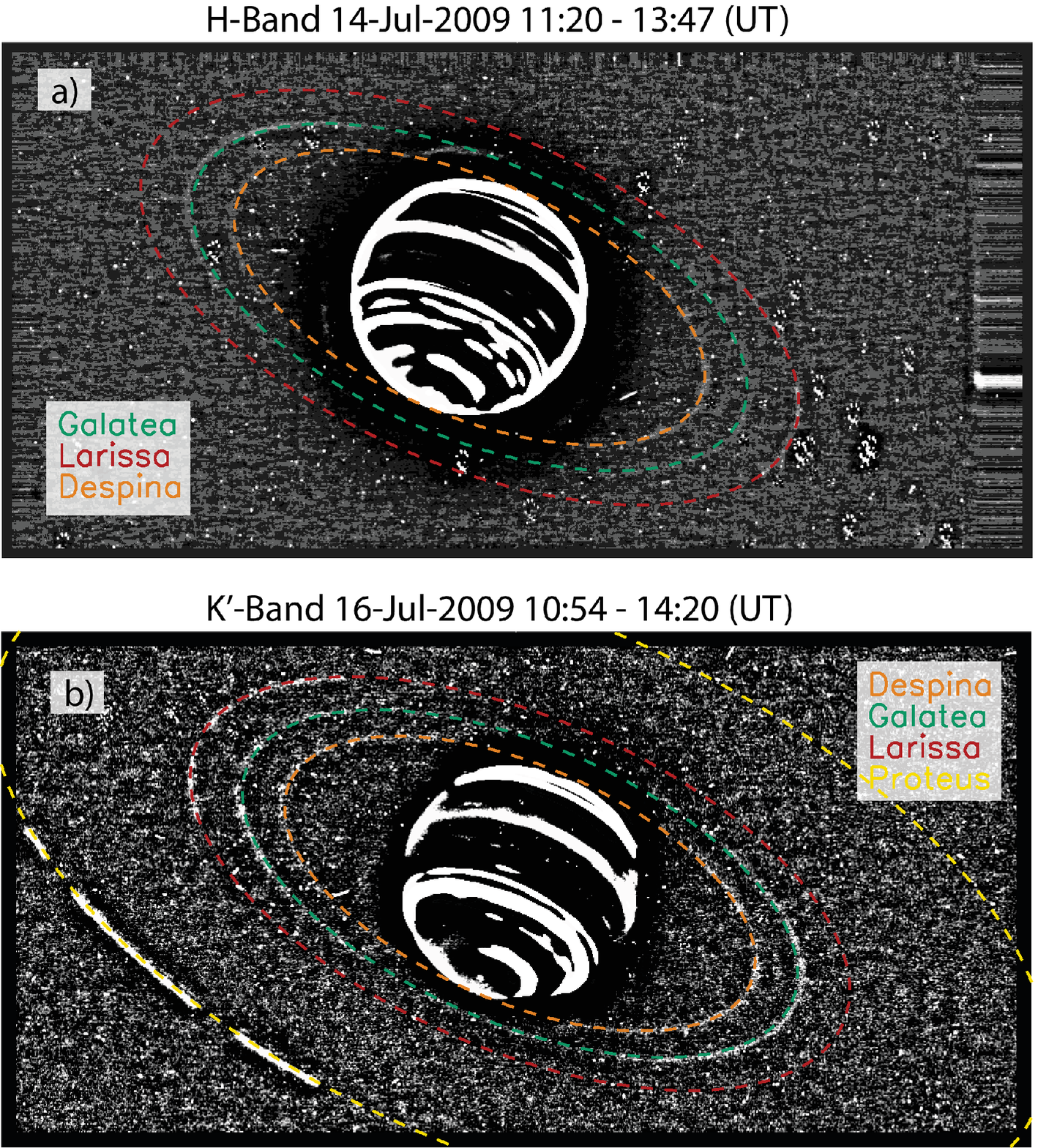}
\caption{Mean averaged, high-pass filtered H- (panel
  \textbf{a}) and K'-band (panel \textbf{b}) images.  H-band image is averaged from the full stack of aligned
    images from 2009 14 July 11:20:21 - 13:47:44 (UT), and K'-band image is averaged
    from the stack of aligned images from 2009 16 July 10:54:57 - 14:20:09 (UT).
    \textbf{Color dashed lines} indicate the derived apparent orbits of Galatea
    (\textbf{green}), Larissa (\textbf{red}), and Despina (\textbf{orange}) in each image.  In
    K'-band Proteus (\textbf{yellow}) is also indicated.  These align
    well with the true traced moon orbits in each image.}
\label{average}
\end{figure*}

The accuracy of the alignment of our images can be judged from Figure \ref{average}, which shows mean averages of the stack of aligned images in each filter.  The positions of each individual moon trace out
visible orbits in the averaged image.  In order to better resolve these orbits we high-pass filter each averaged image by subtracting from it an identical image that has been median-smoothed with a width of 30
pixels.  High-pass filtering the averaged image
eliminates large features that dominate the intensity range and allows
high resolution of faint structure such as Neptune's moons
and rings.  Neptune's Adams and Le
Verrier rings are clearly visible in both images in Figure
\ref{average}.  In both H- and K'-bands we can clearly distinguish the
orbits of Despina, Galatea, and Larissa, although those of Despina and
Galatea are difficult to distinguish from the nearby Le Verrier and
Adams rings.  In K'-band we can also distinguish the orbit of Proteus.
We superpose the orbits we derived for these moons in Figure
\ref{average}.  The moon orbits
physically traced in our averaged high-pass filtered images align well with the
superposed derived moon orbits, reflecting the accuracy of our alignment
and Neptune center determinations.

\begin{figure*}[t]
\includegraphics[width=1.0\linewidth]{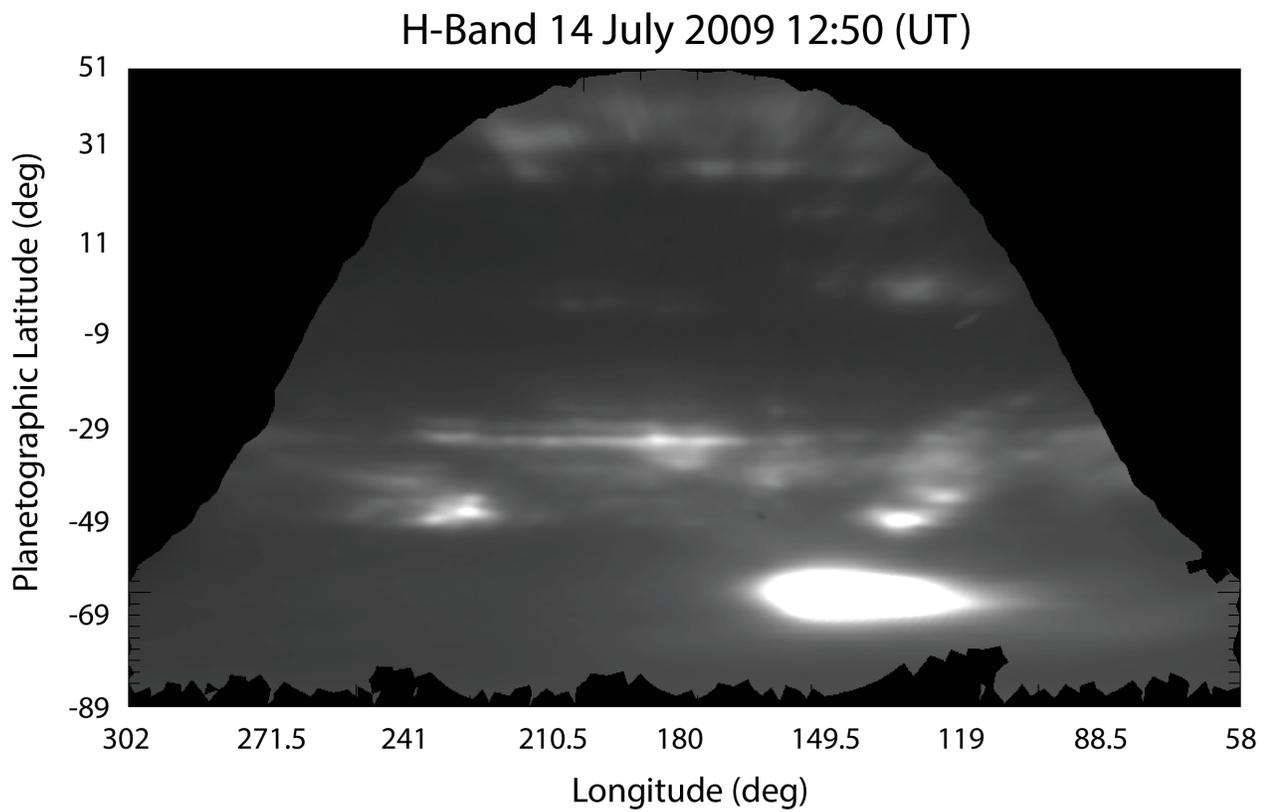}
\caption{Transformed image, taken in H-band on 2009 July 14 12:50:29 (UT).}
\label{trans}
\end{figure*}

After precise location of Neptune's center, we transform the data to regularly gridded, cylindrical 
coordinates, using the same code described in Asay-Davis et al. (2009) and \citet{lii10}.  For 
the transformation from sky coordinates to planetographic
latitude-longitude coordinates on Neptune's 1-bar 
surface, we use equations similar to those in \citet{hueso10}, but simplified with the plane-parallel 
assumption.  Finally, we use IDL's \emph{trigrid} function to 
resample the latitude-longitude data on a regular 
grid.  An example of a transformed image from our cylindrical projection
is shown in Figure \ref{trans}.

\section{Tracking Atmospheric Features}
We use the velocities of cloud features as a tracer for atmospheric wind velocities, and therefore track the
positions of cloud features in our images.  After transformation of each
image frame, we combine the five frames within each sequence together in a mean average.  We track the positions of cloud
features in these transformed averaged images.  On average, zonal drift rates over 5 min are smaller ($<$0.65$^{\circ}$/5 min) than an effective angular resolution element at disc center ($\sim$2.4$^{\circ}$).  Averaging images does not significantly smear features and increases signal-to-noise, allowing us to better distinguish faint, fine features.  Averaging sets of H-band data from 14 July yields 15 averaged
images and averaging sets of K'-band data from 16 July yields 21
averaged images.

\begin{figure*}[t]
\includegraphics[angle=270,width=1.0\linewidth]{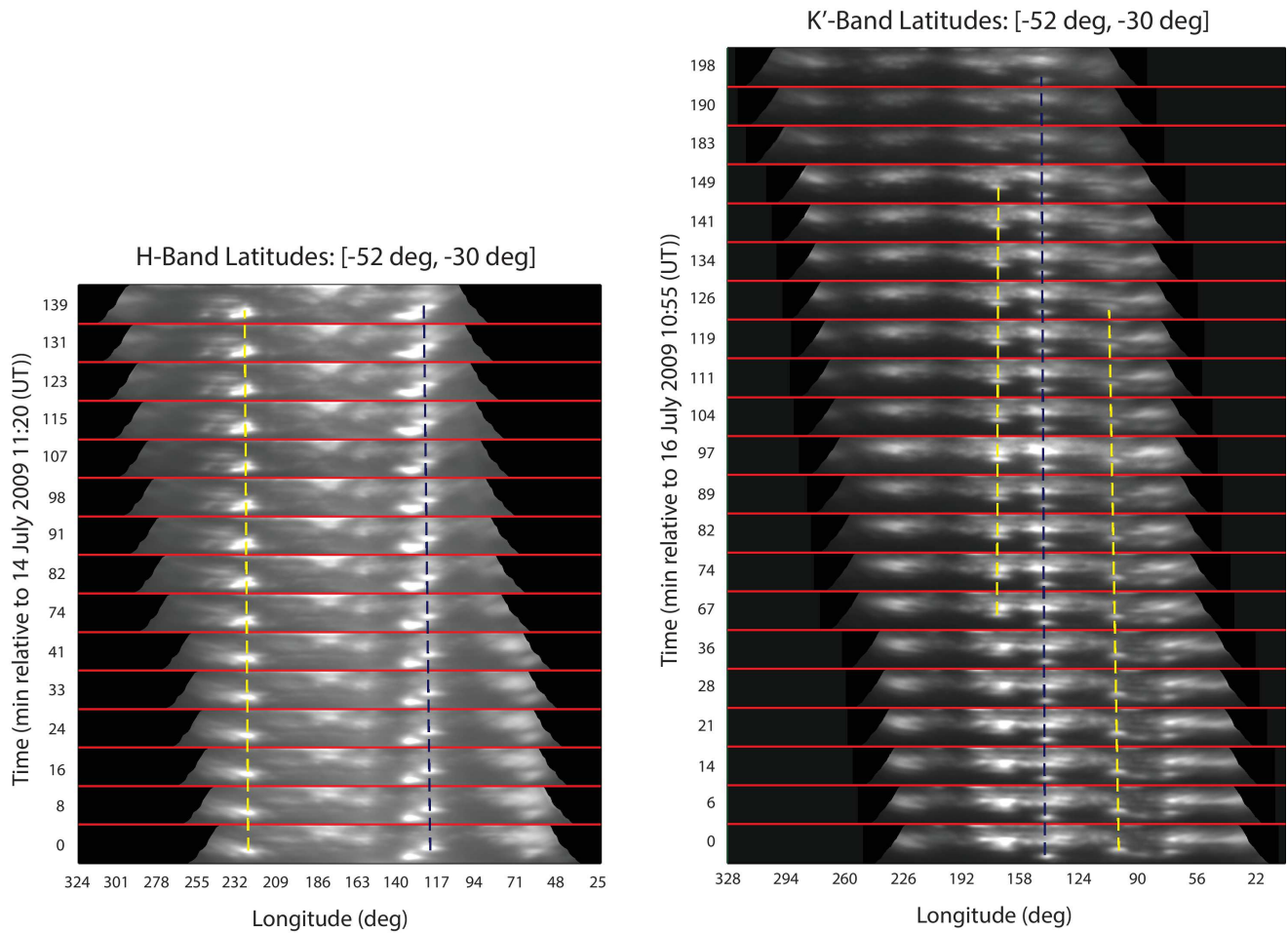}
\caption{Transformed averaged images in strips of fixed latitude
  (52$^{\circ}$S-30$^{\circ}$S) stacked vertically with ascending time.
  Vertical axis indicates the exposure time of the first of the five
  images composing the averaged image.  Repeating \textbf{horizontal
    red} lines mark the boundaries between successive image strips.
  \textbf{Blue dashed} lines indicate representative features which are
  relatively consistent in brightness and morphology while
  \textbf{yellow dashed} lines indicate representative features which are more
  ephemeral and change morphology more significantly and rapidly.}
\label{transithkp}
\end{figure*}

In order to track cloud features we
first identify them between successive transformed averaged images by
constructing images such as Figure \ref{transithkp}, which shows
successive transformed averaged images in strips of fixed latitude (from
52-30$^{\circ}$S) stacked with time increasing along the vertical axis.

We define a feature as having an observed brightness distribution
distinct from features around it and persisting in at least 4 successive
images.  Studying Figure \ref{transithkp}: bright atmospheric features show a broad range of
dynamics.  Some features persist relatively consistently in brightness and
morphology (blue dashed lines), while others are very ephemeral (yellow
dashed lines), appearing and disappearing
or significantly varying in morphology on minute timescales.  It appears that the
smallest features tend also to be the most ephemeral, as was also noted
in the Voyager era \citep{smith89}.  Our initial
interpretation of images such as Figure \ref{transithkp} is that
features in K'-band evolve more rapidly and appear more ephemeral.
This, however, could be due to the greater number of smaller features observed on 16 July.  We compare the dispersion in
zonal velocities between 14 July H- and 16 July K'-band features in Section \ref{s_comp}.

\begin{figure*}[t]
\begin{center}
\includegraphics[scale=0.8]{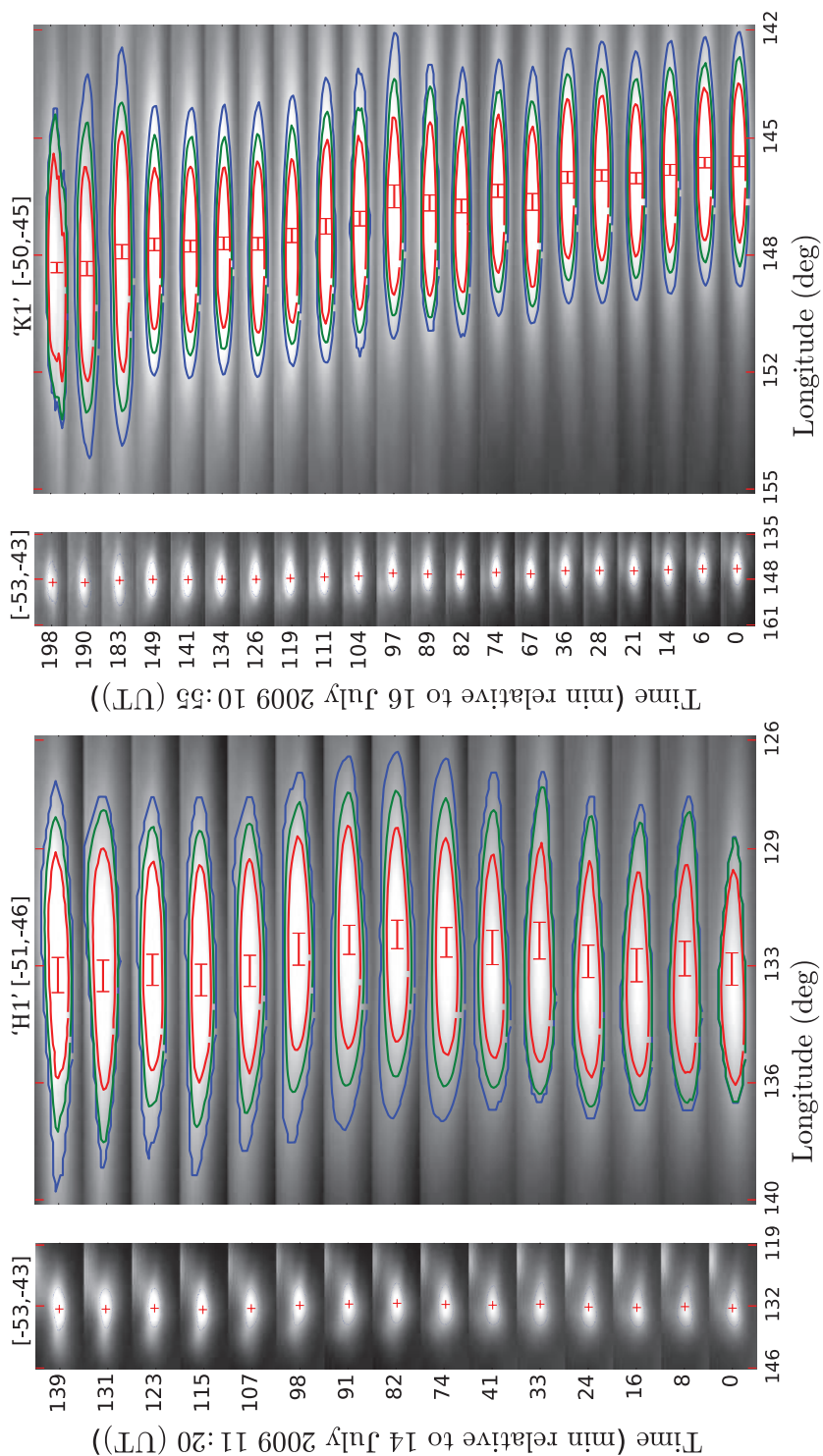}
\caption{Transformed averaged images in strips of fixed latitude -- identified at the top of each column -- stacked vertically with ascending time. Vertical axis indicates the exposure time of the first of the five images composing the averaged image. Second (fourth) column shows a zoomed-in image of the feature shown in the first (third) column, with contour lines at 60$\%$ (\textbf{blue}), 70$\%$ (\textbf{green}), and 80$\%$ (\textbf{red}) the maximum feature intensity, and our derived feature center indicated to illustrate our feature tracking method. Identifying names are shown above the second and fourth columns in order to easily indicate these features in Figures \ref{binh} and \ref{bink}.}
\label{transit_cont}
\end{center}
\end{figure*}

After we identify distinct features, we measure their longitude-latitude positions in each transformed averaged image following the method of \citet{martin12}: for a single feature in a transformed averaged
image we take an initial contour of the image at an
intensity level that outlines that feature.  After isolating
this feature we then take three more contours of this feature at
intensity levels defined at 60$\%$, 70$\%$, and 80$\%$ of the maximum intensity within the initial contour.  We define the center of a feature as the midpoint between longitudinal and latitudinal extrema of a given contour; three individual feature center positions are measured using these three contours.  We define the final feature center position as the average of these
three measured center positions.  The uncertainty in feature center
position is defined as the sum in quadrature of the standard deviation of the three measured center positions and navigation uncertainty associated with location of Neptune center position in untransformed images.  We repeat this procedure for each tracked
feature throughout each of the images in which it is identified. An example illustrating our method is shown in Figure \ref{transit_cont}.

\section{Results}
\subsection{Deriving Zonal Drift Rates}
Our results for 14 July H-band features are shown in Figure \ref{binh} and those for 16 July K'-band features are shown in Figure \ref{bink}.  Here we show the longitude positions of each
tracked feature versus time, separated into latitude bins identified
above each panel.

At fixed latitude, we expect the longitude positions of
atmospheric features to move linearly with time, in agreement with the
smooth Voyager wind profile of \citet{sromovsky93}.  We expect the
latitudinal speeds of most features to be consistent with zero.  We
derive longitudinal and latitudinal drift rates for each feature by
fitting lines to their longitudes vs. time and latitudes vs. time using a Monte-Carlo iteration routine.  Our routine fits position vs. time data to a line using the function $\emph{ladfit}$ in IDL (which fits data to a linear model using a ``robust" least absolute deviation method) with each iteration sampling each position measurement from a normal distribution centered on the original position measurement with a width the size of the uncertainty in that measurement.  We make $\sim$1,000 iterations until
our fits converge.  We use the means of the fit parameters from
all iterations as our output fit parameters, and the standard deviations
of the fit parameters from all iterations as the uncertainties in our
output fit parameters.

Our derived drift rates are shown as solid and dotted
lines over our data in Figures \ref{binh} and \ref{bink}.  Also shown in
Figures \ref{binh} and \ref{bink}, overplotted onto each feature using a dashed line, are slopes indicating
the longitudinal drift rates expected at the latitude of each feature
according to the smooth Voyager wind profile of \citet{sromovsky93}.
We see that, although many features follow constant drift
rates, as we expect, there are also many features whose individual longitude
positions show significant variation off of these constant drift rates.
Non-constant velocities could include true variability, but could also
include measurement errors.  Sources of error include changes in cloud
morphology, image navigation errors, and feature centroiding errors.
For instance, anomalous longitude position measurements at $\sim$70 min on July 14 at multiple latitudes argue strongly for some type of measurement error.  Therefore, in order to
disentangle the effect of large variations in individual longitude position
measurements, either real or due to error, from dispersion in derived mean longitudinal drift rates, we separate features by the mean of their absolute residuals in individual
longitude position about their derived longitudinal drift rates,
$\sigma_{rl}$. Our selection is shown in Figure \ref{histdev}.  We divide features into
``Low-$\sigma_{rl}$'' ($\sigma_{rl}$ $\leq$ 0.7 deg) and ``High-$\sigma_{rl}$'' ($\sigma_{rl}$ $>$ 0.7 deg).  The best-defined feature tracks are in the Low-$\sigma_{rl}$ bin, which contains 25 of 41 July 14 H-band features,
and 29 of 46 July 16 K'-band features.

A combination of true variability and unknown sources of uncertainty manifests itself as deviations from linear motion. In order to obtain upper limits to uncertainties for the derived zonal velocities of Low-$\sigma_{rl}$ features, we assume all scatter in the individual measurements of Low-$\sigma_{rl}$ features about their smooth motion is due to unknown errors, and do the following: for each Low-$\sigma_{rl}$ feature whose value for the reduced-$\chi^{2}$ of observed longitude-time data about its derived zonal drift rate is greater than 1, we solve for an additional contribution to the total uncertainty in longitude position such that the reduced-$\chi^{2}$ is equal to 1, assuming this additional source of uncertainty to contribute equally to each longitude position measurement for a given feature, and to be random and uncorrelated with the sources of uncertainty in longitude position already considered; that is, for each Low-$\sigma_{rl}$ feature with $\chi_{red}^{2} >$1 we solve the following expression for an additional unknown source of uncertainty in longitude position, $\sigma_{u}$:
\begin{center}
$\chi_{red}^{2}$ = $\frac{1}{N - 2} \sum\limits_{i=1}^{N} \frac{\left( \phi_{i} - \phi_{exp,i} \right)^{2}}{ \sigma_{cent,i}^{2} + \sigma_{nav,i}^{2} + \sigma_{u}^{2}}$ = 1,
\end{center}
where $N$ is the total number of longitude-time measurements, $\phi_{i}$
are the measured longitude positions, $\phi_{exp,i}$ are the
corresponding longitude positions expected from a feature's derived
zonal drift rate, $\sigma_{cent,i}$ is the standard deviation of the
three measured feature center positions composing the mean longitude
position (from the 60$\%$, 70$\%$, and 80$\%$ intensity contours), and
$\sigma_{nav,i}$ is the contribution from navigation uncertainty.  Once
we solve for $\sigma_{u}$ this way, we include it as a contribution to
the uncertainty in each longitude position measurement of its
corresponding feature and recompute that feature's zonal drift rate and
its uncertainty in the same way outlined above.  The resulting longitude
position errors and derived drift rates are those shown in Figures
\ref{binh} and \ref{bink}. We make this correction for all Low-$\sigma_{rl}$ features. To quantify the relative
magnitude of $\sigma_{u}$, we compute the ratio of $\sigma_{u}$ to the
initially estimated sources of longitude position measurement
uncertainty ($\eta_{u,i}$ = $\sigma_{u}$/$\left(
\sigma_{cent,i} + \sigma_{nav,i} \right)$) for each feature.  For 14
July H-band features the mean $\eta_{u}$ is 9.4 and for 16 July
K'-band features the mean $\eta_{u}$ is 4.9.  We make the same
correction for errors in latitude position for
all Low-$\sigma_{rl}$ 14 July H-band features and for 25 Low-$\sigma_{rl}$
16 July K'-band features. Defining a similar $\eta_{u,lat}$
for unknown errors in latitude position: for 14 July H-band features the
mean $\eta_{u,lat}$ is 5.9 and for 16 July K'-band features the
mean $\eta_{u,lat}$ is 3.4. All uncertainties in feature velocities and drift rates presented here are derived including the contribution of $\sigma_{u}$ to measurement uncertainty. Because $\sigma_{u}$ includes both unknown sources of measurement error and true variability, the uncertainties we present in feature velocities and drift rates are upper limits to the true uncertainties.

We distinguish between High- (red) and Low-$\sigma_{rl}$ (blue) features in Figures \ref{binh} and \ref{bink}. We note that there is a greater number of High-$\sigma_{rl}$ features found in
K'-band, probably due to the fact that a greater number of
smaller features (which we said tend to be more
ephemeral) were observed on 16 July.  We note that a number of 16 July K'-band
features display a pattern suggestive of east-west temporal oscillation,
similar to what was observed by \citet{martin12}.  However, this
pattern occurs simultaneously at t$\simeq$50-150 min and with similar periods and phases for a number
of features at different locations on the planet.  This suggests
that the observed pattern may be caused by some type of measurement error, as discussed above.  For this
reason we also classify these features as High-$\sigma_{rl}$.  We focus
on Low-$\sigma_{rl}$ features in our analysis of dispersion in wind
speeds.

Comparing the drift
rates expected from the smooth Voyager wind profile with our derived drift
rates in Figures \ref{binh} and \ref{bink} we note deviation from the smooth Voyager wind profile.  Variation in zonal drift rates within individual latitude bins reveals dispersion about the smooth Voyager wind profile.  While large deviation from the smooth Voyager wind profile is more frequently found among
High-$\sigma_{rl}$ features, for which deviation in drift rate is strongly entangled with scatter in individual position measurement, Low-$\sigma_{rl}$
features show significant dispersion as well.  We note, however, that we cannot fully confirm or quantify dispersion in zonal velocities from Figures \ref{binh} and \ref{bink}, because uncertainties in the derived drift rates are not shown.  We better quantify the dispersion we observe in Neptune's zonal wind velocities in what follows.

\subsection{Dispersion in Zonal Wind Velocities}
We translate the drift rates of atmospheric features into wind velocities according to the following relations:

\begin{equation}
V_{lon} = R_{eq}\cos \theta \frac{d \phi}{dt},
\end{equation}
\begin{equation}
V_{lat} = \left(R_{eq}\sin^{2}\theta + R_{pol}\cos^{2}\theta \right)
\frac{d \theta}{dt},
\end{equation}
where $V_{lon}$ and $V_{lat}$ are the zonal and latitudinal velocities of atmospheric features (m/s),
$R_{eq}$=24,766$\times$10$^{3}$ m is Neptune's equatorial radius, $R_{pol}$=24,342$\times$10$^{3}$ m
is Neptune's polar radius, and $d \phi/dt$ and $d \theta/dt$
are derived zonal and latitudinal drift rates (rad/s),
where, for $d \phi/dt$, motion from astronomical east to west
is taken to be the positive direction.  Our derived zonal velocities for 14 July H- and 16 July K'-band Low-$\sigma_{rl}$ features are shown in Figures
\ref{srovwh_h} and \ref{srovwh_kp}, respectively, against the smooth Voyager wind
profile of \citet{sromovsky93} (black
solid line). They are also listed for each feature in Tables \ref{htab} and \ref{kptab}, along with other relevant quantities associated with the motion of each feature.  We plot features such that the
size of the square used to represent distinct features increases linearly with the length of time over which a feature was
tracked.  The longest and shortest tracking times we obtain in H-band
are 139 min and 24 min, respectively, and the longest and shortest tracking
times we obtain in K'-band are 198 min and 21 min, respectively.

Considering both Figures \ref{srovwh_h} and \ref{srovwh_kp} we observe significant deviation from and dispersion about the smooth Voyager
wind profile.  Low-$\sigma_{rl}$ features which scatter most
from the smooth Voyager wind profile are those which were tracked for shorter
time periods.  Increased dispersion is seen at equatorial and southern
mid-latitudes.  On average, greater dispersion is seen among 14 July H-band
features.

Although Low-$\sigma_{rl}$ 16 July K'-band features are on average found
to be consistent with the smooth Voyager wind profile within
uncertainties, we find significant deviation, $\Delta V_{lon,\sigma}$,
for 16 July K'-band features from the smooth Voyager wind profile
(including its width of uncertainty) as high as 290$\pm$77 m/s (at the
equator).\footnote[2]{Two descriptions of the deviation in
    zonal velocities from the smooth Voyager wind profile are used
    throughout. $\Delta V_{lon,\sigma}$, which is presented here, takes
    into account uncertainty in the smooth Voyager wind profile. For a
    feature that is faster than the smooth Voyager wind profile, $\Delta
    V_{lon,\sigma}$ $\equiv$ $V_{lon} - \left( V_{voy} + \sigma_{voy}
    \right)$, where $V_{voy}$ and $\sigma_{voy}$ are the zonal velocity
    and its uncertainty predicted by the smooth Voyager profile at the
    latitude of the feature considered. For a feature that is slower
    than the smooth Voyager profile, $\Delta V_{lon,\sigma}$ $\equiv$
    $V_{voy} - \sigma_{voy} - V_{lon}$.  According to this definition, a
    positive value of $\Delta V_{lon,\sigma}$ greater than its
    uncertainty represents a feature with significant deviation in zonal
    velocity, while a negative value of $\Delta V_{lon,\sigma}$
    represents a feature that is consistent with the smooth Voyager
    profile. The quantity $\Delta V_{lon}$ $\equiv$ $V_{lon} - V_{voy}$
    measures deviation from the smooth Voyager wind profile without
    consideration of uncertainty in the latter. This quantity is used
    mainly when comparing deviation of two features or sets of
    features. These quantities are specified in context. A Graphical
    illustration of these two definitions of deviation is shown in the
    top legend of Figure \ref{srovwh_h}.} Low-$\sigma_{rl}$ H-band features show a significant average absolute deviation of 177$\pm$55 m/s.  Deviation from the smooth Voyager wind profile of 14 July H-band features is found as high as $\sim$500 m/s, although uncertainties in these measurements are large -- as high as 25$\%$ at 23$^{\circ}$S and 50$\%$ at the equator.  We more closely examine the difference in dispersion between 14 July H- and 16 July K'-band features in Section \ref{s_comp}.

Previous studies tracking the motions of Neptune's cloud features have found significant dispersion and deviation in the zonal velocities. \citet{limaye91} found variation as high as $\sim$750 m/s in the zonal velocities of individual features at fixed latitude, with the greatest variation at equatorial and mid-latitudes. When only considering features whose uncertainties in zonal velocity were $<$25 m/s, the authors found the standard deviation of the zonal velocities about the mean, averaged within 1-degree latitude bins, to be as high as $\sim$275 m/s. \citet{sromovsky01b} tracked the motions of bright cloud features in HST data from 1994, 1995, and 1996, and found deviation in zonal velocities from the smooth Voyager profile as high as $\sim$175 m/s for features tracked over time periods ranging from $\sim$1-18 hrs.  \citet{martin12} found deviation in zonal velocities from the smooth Voyager profile as high as $\sim$500 m/s for features tracked over $\sim$1-4 hrs in NIR Keck AO images of Neptune in 2001.  Using the velocities of cloud features as tracers for atmospheric wind velocities, we agree with previous results and observe significant dispersion in Neptune's wind velocities about its mean zonal wind profile.

\subsection{Cloud Feature Pressures From Radiative Transfer Modeling}
We use radiative transfer modeling to estimate the pressures of a selection of 50
features that were
visible in both H- and K'-bands from the beginning and end of both nights. Model
spectra are produced using a 300-layer two-stream radiative transfer code, which is
described in detail in Appendix B of \citet{statia12}.  We adopt the
temperature profile derived by \citet{fletcher10} throughout the atmosphere.  We
assume a mixing ratio of 0.15 for He and 0.003 for N$_{2}$ \citep{conrath93}. The methane (CH$_{4}$) abundance
follows \citet{fletcher10} in the upper atmosphere and remains at a mole fraction
of 0.022 in the troposphere below the condensation level \citep{baines95}. The gas opacity at these
wavelengths is dominated by H$_{2}$ collision-induced absorption (CIA) and CH$_{4}$ opacity;
for CIA we use the coefficients for hydrogen, helium and methane from \citet{borysow85,borysow88} and \citet{borysow91, borysow92, borysow93}, assuming an equilibrium ortho/para ratio
for H$_{2}$. For methane, we use the correlated-k method and follow the recommendations
of \citet{sromovsky12} for outer planet NIR spectra.

As the spectral information in our data is limited, we use a simplified model of the
atmospheric cloud/haze distribution. We assume the presence of an optically thick
bottom `surface' cloud and set the depth of this cloud to 2.4 bar. We set the
Henyey-Greenstein asymmetry parameter of the bottom cloud to -0.1 (preferentially
backscattering) and adjust the single scattering albedo to match the data as
described in \citet{statia10}.  Luszcz-Cook (2012) shows that this model
fits the $\mu$ dependence of the spectrum in a dark part of Neptune's
disk in field-integral spectroscopic (OSIRIS) data obtained at Keck.  For this simplified model, we allow for one
additional aerosol layer above the bottom cloud; the particles in this
higher-altitude haze/cloud layer are treated as Mie scatterers: we assume that
ensembles of particles are distributed according to
\begin{equation}
n \left(r \right) \propto r^{6} \exp \left(-6r/r_{max} \right)
\end{equation}
where $n\left( r \right)$ is the number density of particles of radius $r$, and $r_{max}$ is the maximum
in the particle distribution \citep{hansen70} and is set to 1.0 $\mu$m (e.g. \citealt{irwin11}). This particle size distribution is analogous to that found on Titan \citep{mitchell11}.  The
extinction cross section and Henyey-Greenstein asymmetry parameter of the scattering
are calculated using Mie theory. We assume that the particles have a scale height
that is 0.1 times the gas scale height, corresponding to physically thin cloud layers \citep{irwin11}. The free parameters in the model are the
number density of cloud particles at the bottom (maximum) pressure of the
haze/cloud, and the altitude (pressure) of the cloud. For each feature identified in
the data for modeling, we fit the observed H- and K'- band I/F values given the
viewing geometry ($\mu$) in the following way: for each of 40 model pressure levels
distributed (logarithmically) from 3 mbar to 3 bar, we determine the cloud particle
number density that would best match the model H-band I/F to the observed H-band
I/F. We repeat this procedure for K'-band, then we find the pressure at which the H
and K'-band best-fit number densities agree; that is, we determine the pressure at
which a cloud of some particle density can best reproduce both the H- and K'-band
I/F values.

Cloud-top pressure retrievals are shown in Figures \ref{deep}$\mathrm{a}$ and \ref{deep}$\mathrm{b}$.  At different latitudes, the range of derived cloud pressures varies by as much as a factor of $\sim$4, with the greatest range in pressures found at southern mid-latitudes (60$^{\circ}$S to 27$^{\circ}$S), spanning $\sim$0.5 bar ($\sim$2 scale heights).  Clouds at northern
latitudes (above 20$^{\circ}$N) lie at higher altitudes
($\sim$0.1-0.2 bar).  We note, however, that clouds at northern
latitudes are systematically higher in emission angle, and from these
limited data we cannot rule out a systematic bias in the model with
$\mu$.  Although the 2D relation of pressure vs. $\mu$ in Figure \ref{deep}$\mathrm{b}$ seems suggestive of a bias with $\mu$, this effect is entangled with any dependence of cloud pressure on latitude, and within
fixed latitude bins (including the 3rd dimension shown by symbol shape
and color in Figure \ref{deep}$\mathrm{b}$) there is not an obvious bias with $\mu$.  For example, aside from one feature at $\sim$50$^{\circ}$S found at $\sim$0.6 bar, clouds at equatorial latitudes
(27$^{\circ}$S to 5$^{\circ}$N) are uniformly found
at deeper altitudes ($\sim$0.5 bar) independent of viewing angle.

Our observation that the northern features appear to be at the highest
altitudes is consistent with previous authors (e.g. \citealt{sromovsky01b, gibbard03, statia12}).  The precise values of the derived northern cloud pressures do not, however, agree with these works, which find them in the stratosphere at 0.023-0.064 bar.  We expect these differences to be related to the limitations of our data (we only have broadband measurements, not spectra), differences in the sensitivity of our measurements to different altitudes (for example, \citet{gibbard03} measures mostly in K'-band, which is not sensitive to altitudes as deep as those we probe in H-band), and the simplicity of our model -- previous studies have favored models with a more complicated haze structure and which vary other model parameters (e.g. \citealt{baines90, gibbard02, irwin11, karkoschka11, statia12}).  Our finding that equatorial
features are deepest, while the SPFs in the south are found above them
($\sim$0.3 bar), is different from the results of \citet{gibbard03}, which suggest a trend of increasing altitude with latitude
from south to north.  However, clouds at these equatorial latitudes were not observed at the earlier epoch (see \citet{gibbard03}).

\subsection{South Polar Features}\label{SPFs}
We observe one large bright SPF at 64$^{\circ}$S on 14 July and three
smaller SPFs centered at $\sim$65$^{\circ}$S on 16 July.  SPFs near these
latitudes have been seen since the Voyager era (e.g. \citealt{smith89, hammel89b, limaye91, sromovsky93, sromovsky01c, rages02, karkoschka11a, martin12}) although they have displayed a latitudinal shift, mostly occurring at 67-74$^{\circ}$S until 2004, and at 60-67$^{\circ}$S since, consistent with our observations \citep{karkoschka11a}.  Individual SPFs are rapidly-evolving features with well-defined
periods that cannot be tracked from one planet rotation to the next, and move through the larger structure of the SPF formation region (e.g. \citealt{smith89, limaye91, sromovsky93}); these features have been observed to form in a fixed longitude region rotating at nearly the planet interior period \citep{hammel89b} -- that inferred from Voyager's radio data tracking Neptune's magnetic field (\citealt{warwick89, lecacheux93}), then to move East with well-defined periods ranging from 11.7 hr at 74$^{\circ}$S to 13 hr at 68$^{\circ}$S, and dissipate before moving halfway around the planet (\citealt{limaye91, sromovsky93}).  At some times, the SPF region can
be completely free of bright cloud features (e.g. \citealt{sromovsky93,
  rages02}).  At some times individual SPFs can sporadically
form a cloud clump that can be the brightest feature on the disk, with
the brightening lasting only tens of hours or less (\citealt{rages02,
  sromovsky01c}), as we observe on 14 July.

At the beginning of
the night on 14 July we observe the large SPF centered at a longitude-latitude position (152$^{\circ}$W, 64$^{\circ}$S) to move with a zonal velocity of
272$\pm$15 m/s, consistent with the smooth Voyager wind profile, and a latitudinal speed consistent with zero.  At the beginning of the night on 16 July we observe three SPFs at centroid positions (130$^{\circ}$W, 67$^{\circ}$S), (119$^{\circ}$W, 63$^{\circ}$S), and (92$^{\circ}$W, 65$^{\circ}$S) all moving with zonal velocities consistent with the smooth Voyager wind profile except for the SPF at 63$^{\circ}$S, which moves slower than the smooth Voyager profile by $\Delta V_{lon,\sigma}$=70$\pm$45 m/s.  Two 16 July SPFs are found with significant north-south velocities: the SPF at 67$^{\circ}$S has a northward velocity of 76$\pm$43 m/s, and the SPF at 63$^{\circ}$S has a southward velocity of 117$\pm$53 m/s.  If we extrapolate the zonal drift rate of the 14 July SPF we expect its centroid location at 10:55 on 16 July (UT) to be at a longitude of 40$\pm$14$^{\circ}$W, whereas the longitudinal extent of the 16 July SPFs does not reach below $\sim$80$^{\circ}$W at that time.  This strongly suggests that the SPFs
observed on the two dates are different features, and that storms
develop and decay on timescales of hours to days, consistent with
previous observations (e.g. \citealt{limaye91, sromovsky93, sromovsky01c, rages02}).

The dynamics underlying the SPFs have not yet been fully addressed.  \citet{sromovsky93} found evidence for strong convection driving the SPFs. Karkoschka (2011a) found a rotational lock between the SPF formation region and the South Polar Wave (SPW), a southern n=1 wave spanning the latitudes 65-55$^{\circ}$S, visible as a dark band in Voyager and HST data (e.g. \citealt{smith89, sromovsky01a, sromovsky01b}). The authors suggested that the vertical motions causing the formation of SPFs are dynamically linked to the SPW, and that the SPW itself is vertically phase-locked with the planet interior.  The authors indeed used the motions of the SPF formation region and SPW to infer Neptune's rotational period (15.9663$\pm$0.0002 hr), different from the 16.11 hr period measured using Voyager's radio data tracking Neptune's magnetic field (\citealt{warwick89, lecacheux93}).  There is still much to be understood about the dynamics of the SPFs, including their temporal variability.

\section{Discussion}
Here we briefly discuss a few mechanisms that can cause dispersion in Neptune's zonal winds which can be addressed by our observations.  We then compare the dispersion observed in H-band on 14 July with that observed in K'-band on 16 July.

\subsection{Sources of Dispersion}\label{s_source}
We can constrain the contribution of
vertical wind shear to the zonal wind dispersion we observe using our results for the
cloud-top pressures of features.  We find a range in pressures
from $\sim$0.6 bar (at 50$^{\circ}$S) to $\sim$0.1 bar (northern features),
extending approximately 2 scale heights.  Voyager IRIS observations of temperature as a function of latitude suggest that at these pressures (30-1000 mbar)
Neptune's vertical wind shear can be on the order of
30 m/s per scale height, with a maximum near the equator \citep{conrath89}.  For a range in altitudes of 2 scale heights, and assuming that the altitudes of observed cloud features are similar to those for which we derive the cloud-top pressures, vertical wind shear should not contribute more than $\sim$60 m/s to zonal wind dispersion.  If we extend this range of depths up to the 2.4 bar bottom cloud we assume in our models and down to 0.02 bar
(lower limit for northern features found by \citet{gibbard03}), then feature altitudes span no more than $\sim$5 scale heights.  For this range of depths
vertical wind shear should not contribute more than $\sim$150 m/s to
zonal wind dispersion.  In many cases we see greater dispersion in zonal wind velocities than what is expected from vertical wind shear,
even when considering only Low-$\sigma_{rl}$ features, and especially in
H-band.  Vertical wind shear cannot be the only cause of the dispersion we observe in the zonal velocities.

We can more directly limit the relative contribution of vertical wind
shear to dispersion we observe in the zonal velocities: Figure \ref{deep}$\mathrm{c}$ shows the cloud-top pressures vs. residuals in zonal velocity
from the smooth Voyager wind profile, $\Delta V_{lon}$, of Low-$\sigma_{rl}$ features which were visible in both H- and K'-bands towards the beginning of each night, separated into thin
latitude bins extending from 32$^{\circ}$S to 28$^{\circ}$N -- where, at
these altitudes, vertical wind shear is expected to be most important (see
\citet{conrath89}).  Although we do not observe features in any single
latitude bin over a wide enough range of pressures to see the vertical
wind shear clearly manifest itself given our uncertainties, Figure \ref{deep}$\mathrm{c}$ directly indicates
that one or more other mechanisms dominate over vertical wind shear in producing
dispersion in the zonal velocities: there is significant zonal dispersion within latitude bins
where features are all estimated to be at about the same pressure
(1$^{\circ}$S to 4$^{\circ}$N and 26-28$^{\circ}$N).  The zonal dispersion at fixed pressure and at fixed latitudes between 1$^{\circ}$S and 4$^{\circ}$N is greater than even the overall upper limit to zonal wind dispersion we expect from vertical wind shear.  Features found at 32-29$^{\circ}$S display the only
significant range of pressures ($\sim$0.25-0.45 bar).  In this latitude
bin we see a spread in zonal velocities significantly greater than what
is expected over the range from vertical wind shear ($<$20 m/s), and we
do not observe a trend towards higher speed with depth (which at these
latitudes is toward more negative velocities), as would be expected from
vertical wind shear (see \citealt{conrath89,conrath91}).

Along with significant variation in zonal wind velocities at fixed latitude, we also observe some significant north-south
feature velocities. Figure \ref{latspeeds} shows the derived north-south
velocities of Low-$\sigma_{rl}$
features against their latitude positions for both 14 July H- (panel
$\mathrm{a}$) and 16 July K'-band (panel $\mathrm{b}$) features.  For a few Low-$\sigma_{rl}$ features we find north-south
velocities that are convincingly significant, and as high as $\sim$200 m/s.  We find that a few features which
display significant north-south velocities also display significant deviation in zonal velocities from the smooth Voyager wind
profile.  Noting significant examples: the 14 July H-band feature at the
equator with residual zonal speed $V_{lon,\sigma}$=480$\pm$254 m/s has a southward wind speed
215$\pm$85 m/s.  A 14 July H-band feature at 7$^{\circ}$N is found with residual zonal speed 180$\pm$24 m/s and northward speed 148$\pm$15 m/s.  We find a 16 July K'-band feature at
37$^{\circ}$S with a northward wind speed of 180$\pm$45 m/s and a residual zonal speed of 117$\pm$80 m/s.  Again we note the 16 July SPF at 63$^{\circ}$S with southward velocity 117$\pm$53 m/s and zonal speed residual 70$\pm$45 m/s.  Significant north-south velocities
and deviations in zonal velocity from the mean zonal profile indicate the presence of one or more
mechanisms which can cause both north-south and east-west residual motions.  In
particular, although we do not confirm the presence of temporal
oscillation in our observations, we cannot rule out vortices causing the dispersion we observe.  Cloud features centered on
or with motions near a vortex would display significant
north-south and east-west temporal oscillations, and if these had
sufficiently long periods the motions would appear linear on these
timescales, with a slope different from the smooth Voyager profile, as we
observe.  Vortices have often been associated with the dark spots
observed on Neptune in Voyager images (such as Voyager GDS and DS2; e.g. \citealt{polvani90, sromovsky93}) and in HST images (such as
NGDS-32; e.g. \citealt{hammel95, sromovsky01b,
  sromovsky02}).

The observed dispersion in zonal
velocities and significant north-south velocities imply that we cannot
rule out the contribution of wave mechanisms with sufficiently long periods, such as Rossby
waves.  We do not observe the presence of wave mechanisms which
oscillate with periods on the order of our observing period.  In particular, we
do not observe evidence of tidal forcing by Triton, as was suggested by
\citet{martin12}.  Even if we were to assume the motions of those
features we noted seemed suggestive of east-west temporal oscillation to
be real and not due to error, their periods would not be near the M2
period of tidal forcing by Triton of 7.24 hrs.

From
ground-based spectroscopic observations prior to the Voyager encounter,
\citet{baines90} found evidence that dynamically driven sublimation and
condensation resulting from vertical motions in the atmosphere are
responsible for rapid changes observed in Neptune's clouds.
\citet{limaye91} also considered this in the context of the
dispersion they observed in the zonal velocities of features in Voyager
images.  The analysis of \citet{martin12} supported this conclusion.
The authors noted that if feature velocities represent true fluid
velocities, for the fluid to remain nearly divergence-free (for
sub-sonic flow), the variation they observed in the zonal velocities
with east-west distance ($>$400 m/s over $\sim$20,000 km) would imply
large east-west gradients in north-south velocities, which they did not
observe, or large vertical motions incompatible with the
atmosphere's gradient Richardson number at the observed altitudes.  They
therefore confirmed that at least some of the dispersion in zonal
velocities they observed is due to transient clouds which do not move
with the flow.  Our observations also support this result.  The spatial
distribution of variation in velocities we observe implies that at least some
of the dispersion is due to transient clouds, and evidence of cloud
motions due to dynamically driven sublimation and condensation is
provided by our observations of clouds that are very ephemeral and
rapidly change morphology.

\subsection{Comparing Dispersion}\label{s_comp}
The results shown in Figures \ref{srovwh_h} and \ref{srovwh_kp} are
immediately suggestive of an overall difference in the magnitude of
dispersion in zonal wind velocities about the smooth Voyager wind profile
between features tracked in H- and K'-bands on 14 and 16 July,
respectively. Whereas 16 July K'-band features seem to agree reasonably well
with the smooth Voyager wind profile, 14 July H-band features appear to show
overall greater dispersion and deviation from the smooth Voyager profile.  The difference in zonal dispersion and deviation between H- and K'-bands might be expected to be a result of the greater range in altitudes probed in H-band over K'-band. Features in H-band can be
seen down to larger depths than those in K'-band because the strong
absorption of H$_{2}$ and CH$_{4}$ in K'-band limits detection of
features only to higher altitudes (see \citet{statia10} Fig. 4).  If the difference in zonal
velocities is due to the difference in the range of depths probed in H-
and K'-bands, then the likely responsible mechanism is vertical wind
shear.  The effects of vertical wind shear in this difference might be seen in two ways: first, we would expect to see a greater spread in the zonal velocities of features tracked in H-band at fixed latitude; second, and perhaps more subtly, we might
expect to see average shifts in the zonal velocities of features between
the two bands, consistent with features tracked in H-band on average
being at greater depths than those tracked in K'-band.  To test the latter of these possibilities we fit
Low-$\sigma_{rl}$ $V_{lon}$-$\phi$ data to a polynomial function of the
form $V_{lon}$ = $a$ + $b\phi^{2}$ + $c\phi^{4}$
to obtain the best-fit values of $a$, $b$, and $c$ using a
Levenberg-Marquardt method, for comparison with the smooth Voyager wind
profile fit of \citet{sromovsky93}.\footnote[3]{We note that Voyager and HST observations were mainly at visible wavelengths and are probably sensing more deeply than K'-band images.}  The results of our polynomial
fits and the 1$\sigma$ uncertainties (calculated from the variance modified by a factor of the reduced-$\chi^{2}$) are shown with solid and dashed red lines in Figures \ref{srovwh_h} and \ref{srovwh_kp} and are listed in Table \ref{poly}, along with the smooth Voyager profile fit of \citet{sromovsky93}.  The zonal velocities of Low-$\sigma_{rl}$ features measured
in K'-band on 16 July agree well with the smooth Voyager wind
profile (within 1$\sigma$), whereas the zonal velocities of Low-$\sigma_{rl}$
features measured in H-band on 14 July are best represented by a
profile which is shifted towards more positive velocities by 180$\pm$50
m/s.

\begin{table*}
\small
\begin{tabular}{@{}llll@{}}
\tableline
Data & Constant Term (m/s) & $\phi^{2}$ Term (m/s/deg$^{2}$) &
$\phi^{4}$ Term (m/s/deg$^{4}$)\\
\tableline
Voyager & -398$\pm$12 & 0.188$\pm$0.014 & -1.2E-5$\pm$0.3E-5\\
14 July H-Band & -205$\pm$49 & 0.083$\pm$0.06 & 7.0E-6$\pm$14.0E-6\\
16 July K'-Band & -387$\pm$37 & 0.165$\pm$0.041 & -1.3E-5$\pm$1.1E-5\\
\tableline
\end{tabular}
\caption{Polynomial fits to feature zonal velocities vs. latitudes and
  their 1$\sigma$ uncertainties. }
\label{poly}
\end{table*}


Interestingly, this shift is in the opposite direction from what we would expect due to vertical wind shear.  If we assume that most features shown in Figures \ref{srovwh_h} and \ref{srovwh_kp} are at
similar altitudes as those for which we measure the cloud-top pressures, or at least that the zonal direction of the vertical wind shear with pressure is the same at the altitudes of these features as those to which the results of \citet{conrath89} and \citet{conrath91} apply, then as a result of vertical wind shear we would expect a general decay in zonal speed with height.  We observe a shift of our 14 July H-band profile fit towards lower zonal speeds (more positive velocities) at latitudes from $\sim$30$^{\circ}$S-30$^{\circ}$N, where the results of \citet{conrath89} and \citet{conrath91} show the vertical wind shear is most important, and where features that most strongly drive the offset in our profile fit are found.  The shift in our profile fit is in the direction we would expect if 14 July H-band features were on average found at higher altitudes than K'-band features.  The effects of vertical wind shear are not found in differences in the average zonal velocities of 14 July H- and 16 July K'-band features.  We note for clarity that we do not interpret differences in our profile fits to the latitudinal distribution of zonal velocities of Low-$\sigma_{rl}$ features as evidence for a consistent change in Neptune's mean zonal wind profile from the smooth Voyager wind profile -- the short observing times and latitude gaps in our data make
it ill-suited to yield evidence for such a claim.  In what follows we continue to interpret the smooth Voyager wind profile as describing Neptune's mean zonal wind profile.

If vertical wind shear were the cause
of the apparent difference in zonal dispersion and deviation between 14 July H- and 16 July
K'-band features then we would also expect: 1) greater spread in 14 July
H-band feature zonal velocities over that of 16 July K'-band feature
zonal velocities at fixed latitude; 2) greater deviation from the smooth
Voyager wind profile of 14 July H-band feature zonal velocities over
that of 16 July K'-band feature zonal velocities at fixed latitude (and this difference in deviation would be fully accounted for by the difference in spread in zonal velocities); 3) differences in the deviation and spread in zonal velocities would not exceed the upper limits we expect for dispersion in the zonal velocities due to vertical wind shear (Section \ref{s_source}).  To explore this
we do the following: we separate $\Delta V_{lon}$ for 14 July H- (squares) and 16 July
K'-band (triangles) Low-$\sigma_{rl}$ features (shown in Figure \ref{pans}$\mathrm{a}$) into
latitude bins (shaded bars) chosen according to the
latitude bands along which features are observed to be centered (see Figure
\ref{discs}).  In each of these latitude bins we compute a weighted
average of the absolute value of residuals in zonal velocity of Low-$\sigma_{rl}$ features from the smooth Voyager wind profile, mean $\left| \Delta V_{lon} \right|$, comparing features tracked in H- and
K'-bands on 14 and 16 July, respectively (we do not consider uncertainty in the smooth Voyager profile in this calculation, and so these results are useful mainly as a comparison between 14 July H- and 16 July K'-band features).  The results are shown in Figure \ref{pans}$\mathrm{b}$.  On average, 14 July H-band features show greater deviation from the smooth Voyager wind profile than K'-band features,
and especially at equatorial and southern
mid-latitudes.

To directly examine differences in the dispersion in zonal velocity, within each
latitude bin where two or more Low-$\sigma_{rl}$ features are found, we
compute the difference between the fastest and slowest Low-$\sigma_{rl}$ zonal velocities, max $\delta V_{lon}$ $\equiv$ $\left( \mathrm{max} \, V_{lon} - \mathrm{min} \, V_{lon} \right)$,
comparing between 14 July H- and 16 July K'-band features.  This is shown in Figure \ref{pans}$\mathrm{c}$.  In many latitude bins uncertainties in the zonal velocities are too large to determine whether or not differences in mean $\left| \Delta V_{lon} \right|$ between 14 July H- and 16 July K'-band features can be fully accounted for by differences in their max $\delta V_{lon}$.  Only in the bin of features centered at 30$^{\circ}$S is there a significant difference in max $\delta V_{lon}$ that can fully account for a corresponding significant difference in mean $\left| \Delta V_{lon} \right|$.  However, our comparison in Figure \ref{pans} is still useful for making statements about the relative contribution of vertical wind shear to differences in the dispersion in zonal velocities: there are differences in max $\delta V_{lon}$ between Low-$\sigma_{rl}$ 14 July H- and 16 July K'-band features in latitude bins centered at 45$^{\circ}$S and 23$^{\circ}$S that are much greater than overall upper limits to dispersion we expect from vertical wind shear.  These indicate the dominance of one or more mechanisms other than vertical wind shear causing differences in the dispersion in zonal velocities between 14 July H- and 16 July K'-band features at these latitudes.  Maximum possible difference in max $\delta V_{lon}$ in the latitude bin centered at 37$^{\circ}$S cannot account for the difference in mean $\left| \Delta V_{lon} \right|$ in the same latitude bin; the maximum possible increase in max $\delta V_{lon}$ of these 14 July H-band features over that of 16 July K'-band features (calculated at opposite extremes of their uncertainties) is 172 m/s, whereas 14 July H-band features at these latitudes display mean $\left| \Delta V_{lon} \right|$ greater than that observed for 16 July K'-band features by $>$250 m/s.  If we assume that mechanisms which cause deviation in zonal velocities from the smooth Voyager wind profile are similarly capable of causing dispersion in zonal velocities, this also indicates the presence of a mechanism other than vertical wind shear which causes differences in the dispersion in zonal velocities of 14 July H- and 16 July K'-band features.

Our observations indicate the dominance of one or more mechanisms over vertical wind shear in producing the overall dispersion we observe in the zonal winds.  That the difference in deviation and dispersion in zonal velocities observed between 14 July H- and 16 July K'-band features is, at least at most latitudes between $\sim$50-20$^{\circ}$S, not primarily attributable to vertical wind shear further suggests that the mechanisms which dominate dispersion in the zonal winds can cause changes in the magnitude of dispersion and deviation in the zonal winds on timescales of hours to days.  This is consistent with the mechanisms discussed above.

\section{Summary and Conclusions}
In this study we tracked the longitude-latitude positions of Neptune's atmospheric features seen in Keck AO images in H-band on 14 July 2009 and in K'-band on 16 July 2009 over time.  We derived zonal and latitudinal drift rates and velocities for these features.  We also performed radiative transfer modeling for features on both nights which were simultaneously visible in both H- and K'-bands.  The results we find are the following:

\begin{enumerate}

\item We find significant dispersion in the zonal velocities of Neptune's cloud features about the smooth Voyager wind profile of \citet{sromovsky93}, with the largest dispersion seen at equatorial and southern mid-latitudes.  The observed dispersion is much greater than the upper limit we expect due to vertical wind shear.  Considering only
  Low-$\sigma_{rl}$ features: deviation in zonal velocity from the
  smooth Voyager wind profile is found as high as 290$\pm$77 m/s (at the
  equator) among features tracked in K'-band on 16 July, and as high as $\sim$500 m/s among features tracked in H-band on 14 July (although uncertainties for these measurements are as high as 25$\%$ at 23$^{\circ}$S and 50$\%$ at the equator).  Comparison of the zonal velocities and cloud-top pressures
  of features at fixed latitude also directly indicates the dominance of
  one or more mechanisms over vertical wind shear in producing
  dispersion in the zonal winds.  Some features display significant
  north-south velocities (as high as $\sim$200 m/s), and a few of these
  also display significant deviation in zonal velocity from the smooth
  Voyager wind profile.  This suggests that zonal dispersion is partially due to a mechanism which can produce residual motion along both directions, such as vortices and wave mechanisms with sufficiently long periods so as to appear linear on these timescales, such as Rossby waves.  Our observations are consistent with the result that zonal dispersion is caused at least in part by transient clouds due to dynamically driven sublimation and condensation.

\item Radiative transfer modeling indicates that, aside from one feature at 50$^{\circ}$S found at $\sim$0.6 bar, cloud features at equatorial latitudes ($\sim$5$^{\circ}$S-5$^{\circ}$N and 27$^{\circ}$S-15$^{\circ}$S) uniformly lie deeper in the atmosphere ($\sim$0.5 bar) while clouds in the north (above 20$^{\circ}$N) are found at higher altitudes ($\sim$0.1-0.2 bar).  Due to limitations of our data, differences in the sensitivity of our measurements to different altitudes, and the simplicity of our model, the precise altitudes we measure for the northern features are different from the results of previous studies, which place them higher, in the stratosphere.

\item  We observe one large SPF at 64$^{\circ}$S on 14 July and three smaller SPFs centered at about the same latitude on 16 July.  Two 16 July SPFs display significant north-south velocities, and that with the largest north-south speed (117$\pm$53 m/s southward) also shows significant deviation from the smooth Voyager wind profile (70$\pm$45 m/s slower).  Extrapolation of the zonal drift rate of the 14 July SPF and comparison with the average position and spatial extent of the 16 July SPFs shows that the SPFs observed on the two nights are different features, indicating that storms can develop and decay on timescales of hours to days, in agreement with previous observations.

\item There is greater dispersion and deviation observed in the zonal
  velocities of features tracked in H-band on 14 July than in those
  tracked in K'-band on 16 July. Polynomial fits to the zonal
  velocities vs. latitudes of our data show that while 16 July K'-band
  feature zonal velocities agree well with the smooth Voyager wind
  profile, 14 July H-band feature zonal velocities are best described by
  a profile that is shifted toward more positive velocities by
  180$\pm$50 m/s.  This shift is in the opposite direction from what we
  would expect if differences in the deviation in zonal velocities from
  the smooth Voyager wind profile between 14 July H-band and 16 July
  K'-band features were due to vertical wind shear, as a result of
  greater average depth of features tracked in H-band.  Direct
  comparison suggests that the difference in deviation and dispersion in
  zonal velocities between 14 July H- and 16 July K'-band features at
  fixed latitude is not primarily attributable to vertical wind shear.
  This further suggests that mechanisms other than vertical wind shear which dominate dispersion in the zonal winds can cause changes in the magnitude of dispersion and deviation in the zonal winds about the mean zonal wind profile on timescales of hours to days.

\end{enumerate}

\acknowledgements
The authors would like to thank Adam Becker for useful discussions.  This research has been supported in part by the National Science
Foundation Science and Technology Center for Adaptive Optics, managed
by the University of California at Santa Cruz under cooperative
agreement No. AST 9876783, as well as by NSF Grant AST-0908575
to the University of California. The data presented here were obtained
at the W.M. Keck Observatory, which is operated as a scientific partnership
among the California Institute of Technology, the University
of California and the National Aeronautics and Space Administration.
The Observatory was made possible by the generous financial support
of the W.M. Keck Foundation. The authors extend special thanks to
those of Hawaiian ancestry on whose sacred mountain we are privileged
to be guests. Without their generous hospitality, none of the observations
presented would have been possible.

\clearpage

\clearpage

\setcounter{figure}{5}

\begin{figure*}[htbp]
\includegraphics[width=\linewidth]{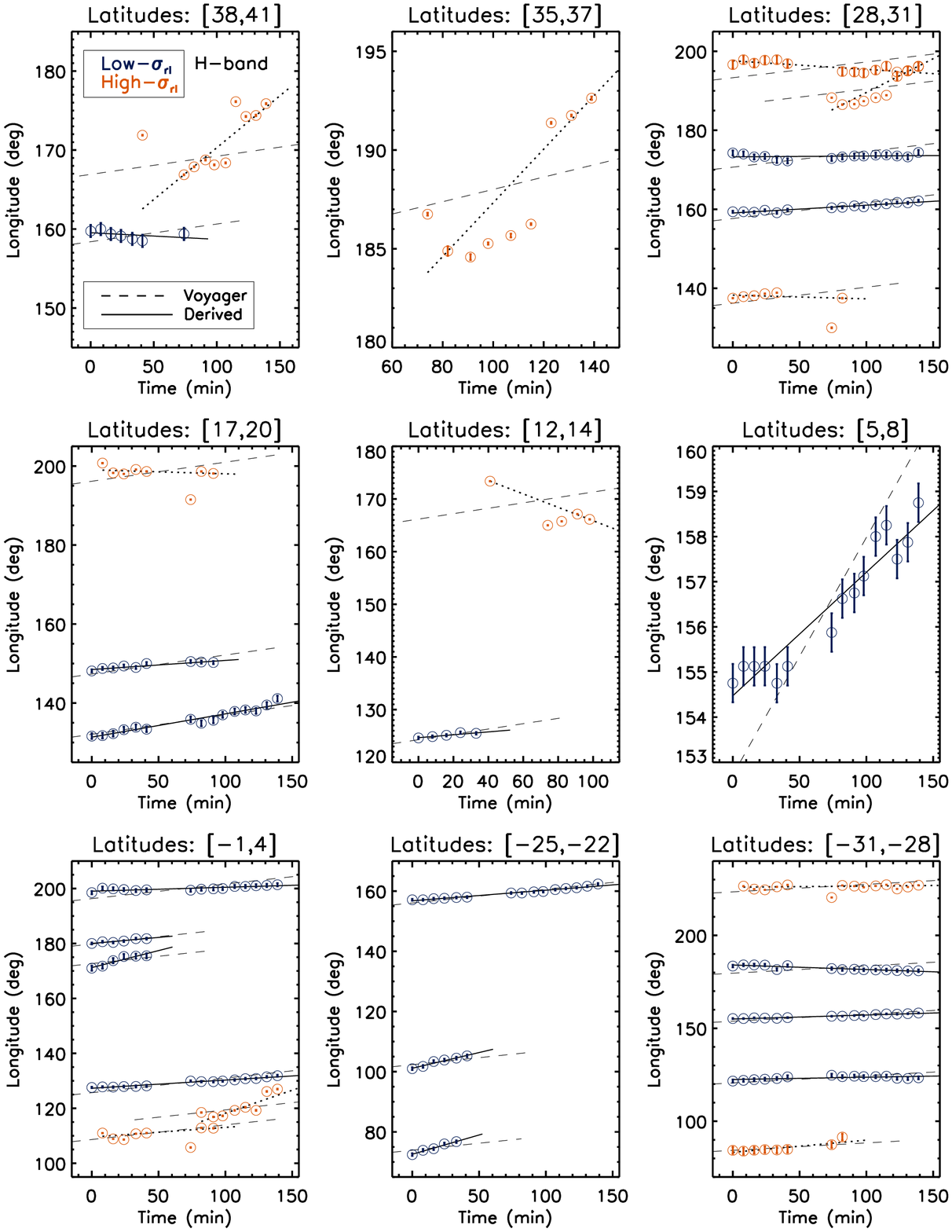}
\caption{}
\end{figure*}

\setcounter{figure}{5}

\begin{figure*}[htbp]
\includegraphics[width=\linewidth]{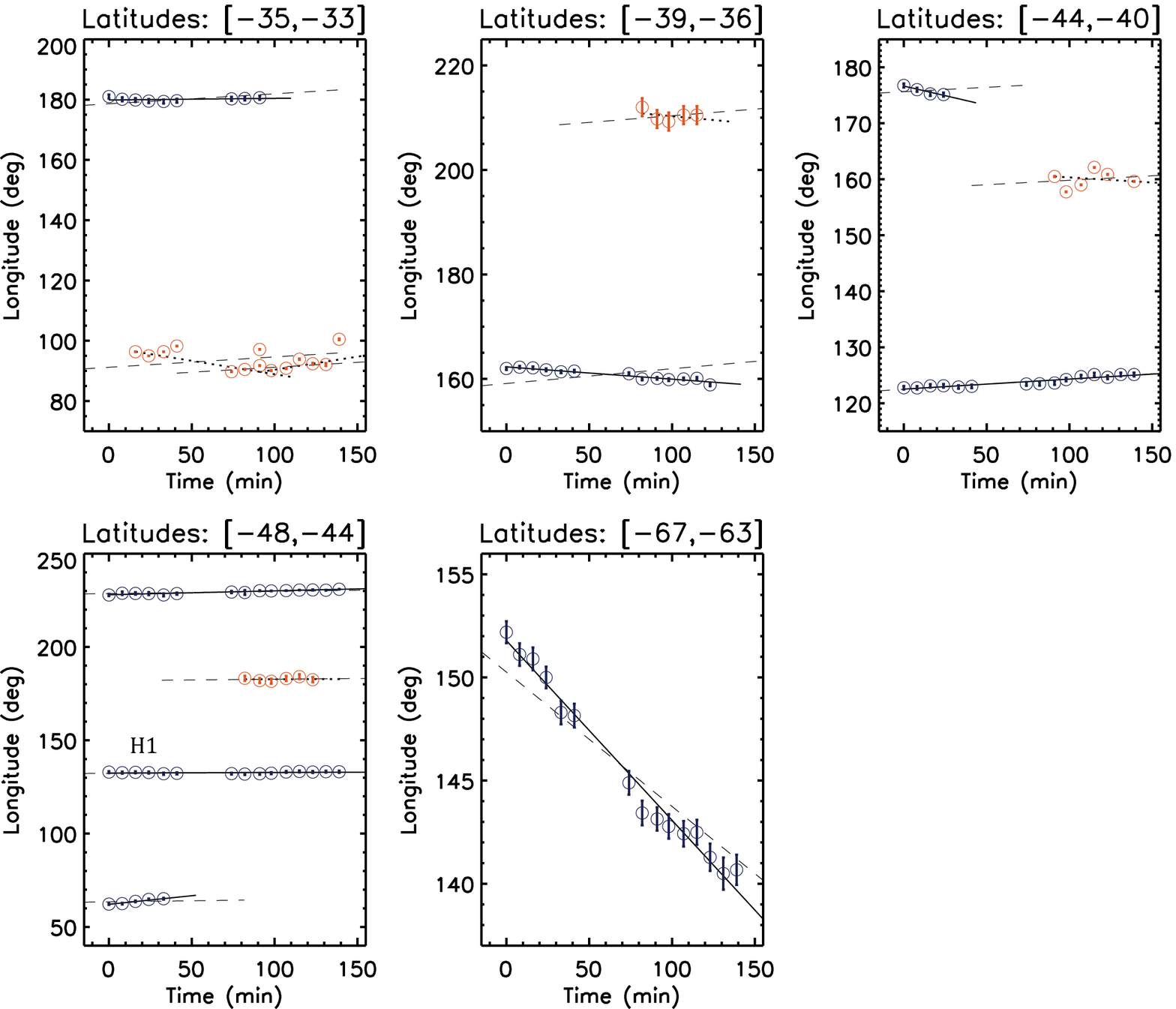}
\caption{Derived zonal drift rates (\textbf{solid lines} and \textbf{dotted lines})
    from the longitude positions versus times of each feature tracked
in H-band images on 14 July, from 11:20-13:47 (UT), separated into latitude bins indicated at
the top of each panel.  Time is measured in minutes relative to 2009 July 14 11:20:20 (UT).  Zonal drift rates expected at the latitude
of each feature according to the smooth Voyager wind profile are shown
for comparison (\textbf{dashed lines}). Features are separated by the
mean of their absolute residuals in measured longitude position about
their derived zonal drift rates: High-$\sigma_{rl}$ (\textbf{red})
and Low-$\sigma_{rl}$ (\textbf{blue}). The feature identified as `H1' in Figure \ref{transit_cont} is indicated. Significant dispersion about the
smooth Voyager wind profile is observed, even for Low-$\sigma_{rl}$
features.}
\label{binh}
\end{figure*}

\setcounter{figure}{6}

\begin{figure*}[htbp]
\includegraphics[width=\linewidth]{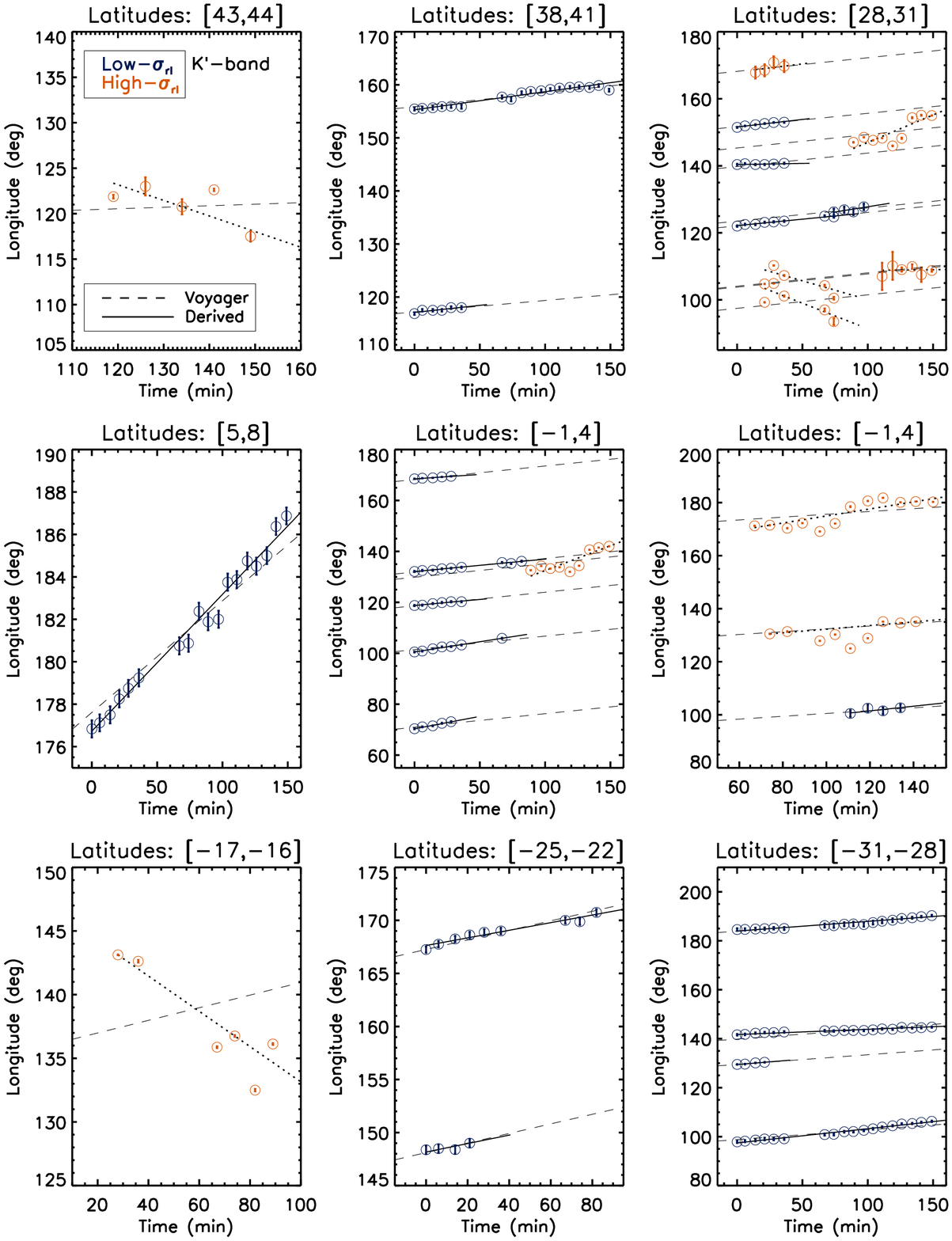}
\caption{}
\end{figure*}

\setcounter{figure}{6}

\begin{figure*}[htbp]
\includegraphics[width=\linewidth]{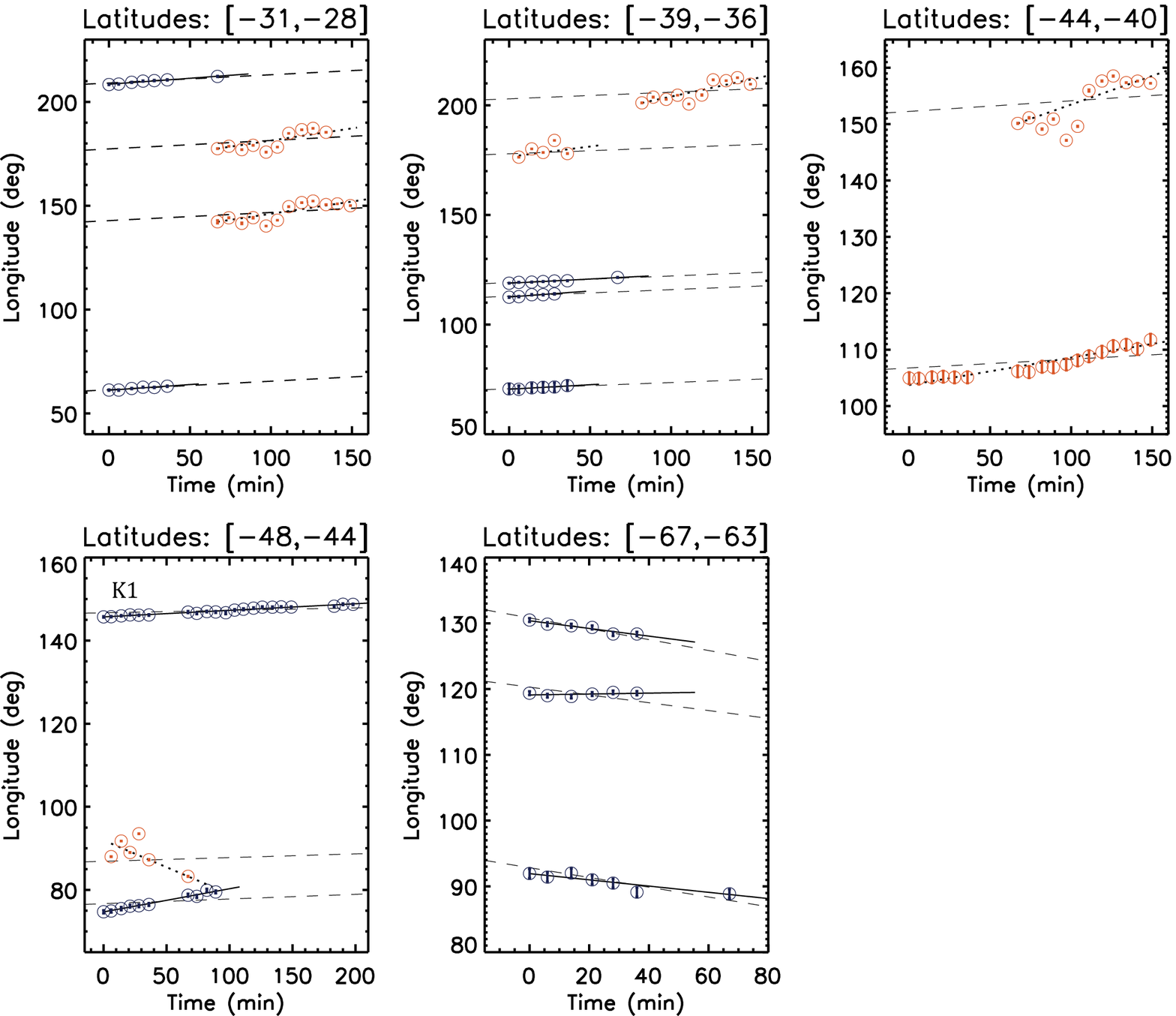} 
\caption{Same as Figure \ref{binh}, now for features tracked in
    K'-band images on 16 July from 10:54-14:20 (UT).  Time is measured
    in minutes relative to 2009 July 16 10:54:57 (UT). The feature
    identified as `K1' in Figure \ref{transit_cont} is
    indicated. Significant dispersion about the smooth Voyager wind
    profile is observed in K'-band on 16 July, even for Low-$\sigma_{rl}$ features.}
\label{bink}
\end{figure*}

\setcounter{figure}{7}
\begin{figure*}[t]
\includegraphics[width=1.0\linewidth]{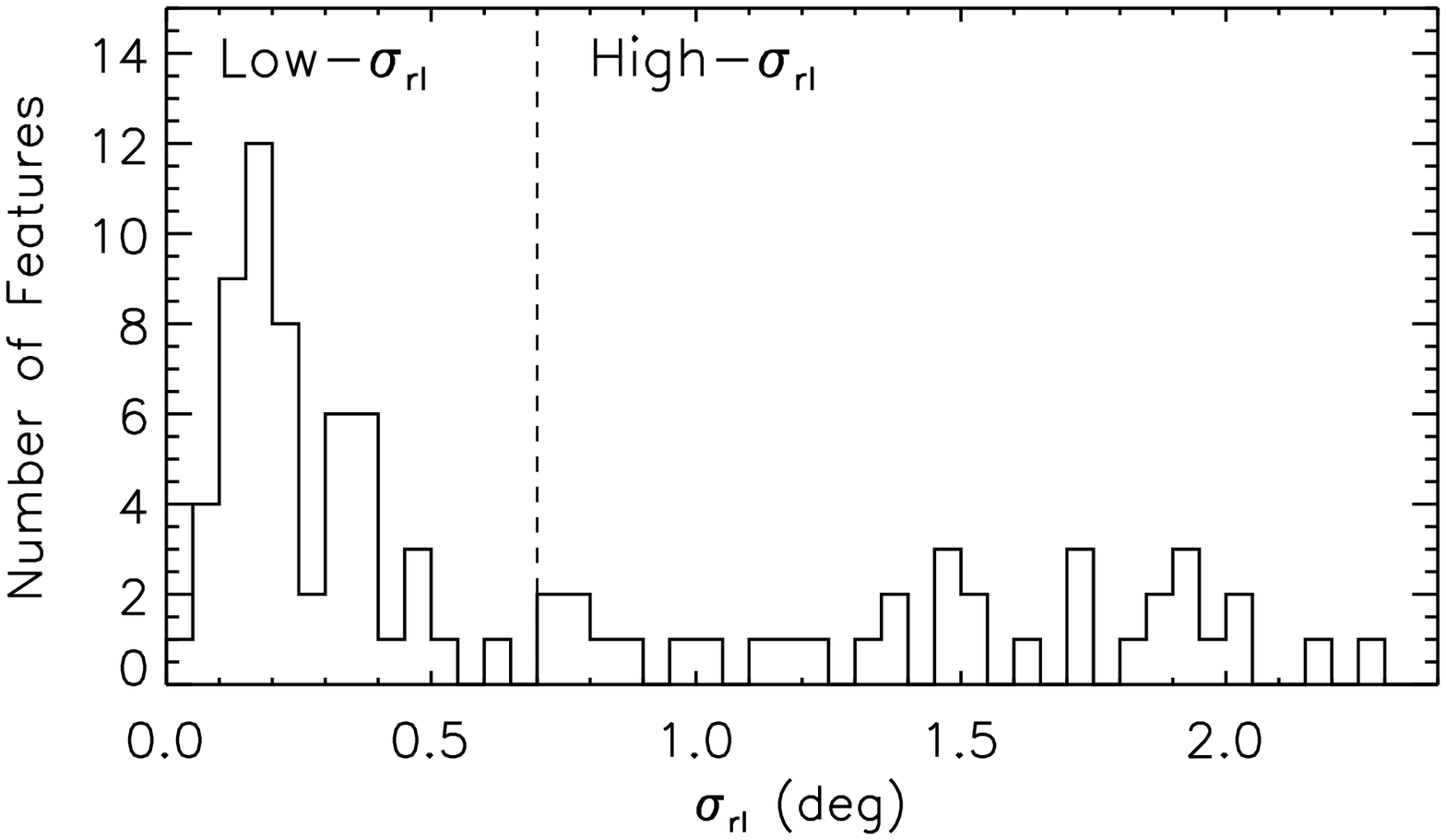}
\caption{Distribution of the mean of absolute residuals in
    measured longitude position about the derived zonal drift rates of
    features, $\sigma_{rl}$.  Our two classifications are shown
    separated by a \textbf{vertical dashed} line: Low-$\sigma_{rl}$ ($\sigma_{rl}$ $\leq$ 0.7 deg) and High-$\sigma_{rl}$ ($\sigma_{rl}$ $>$ 0.7 deg).}
\label{histdev}
\end{figure*}

\begin{figure*}[t]
\begin{center}
\includegraphics[scale=0.7]{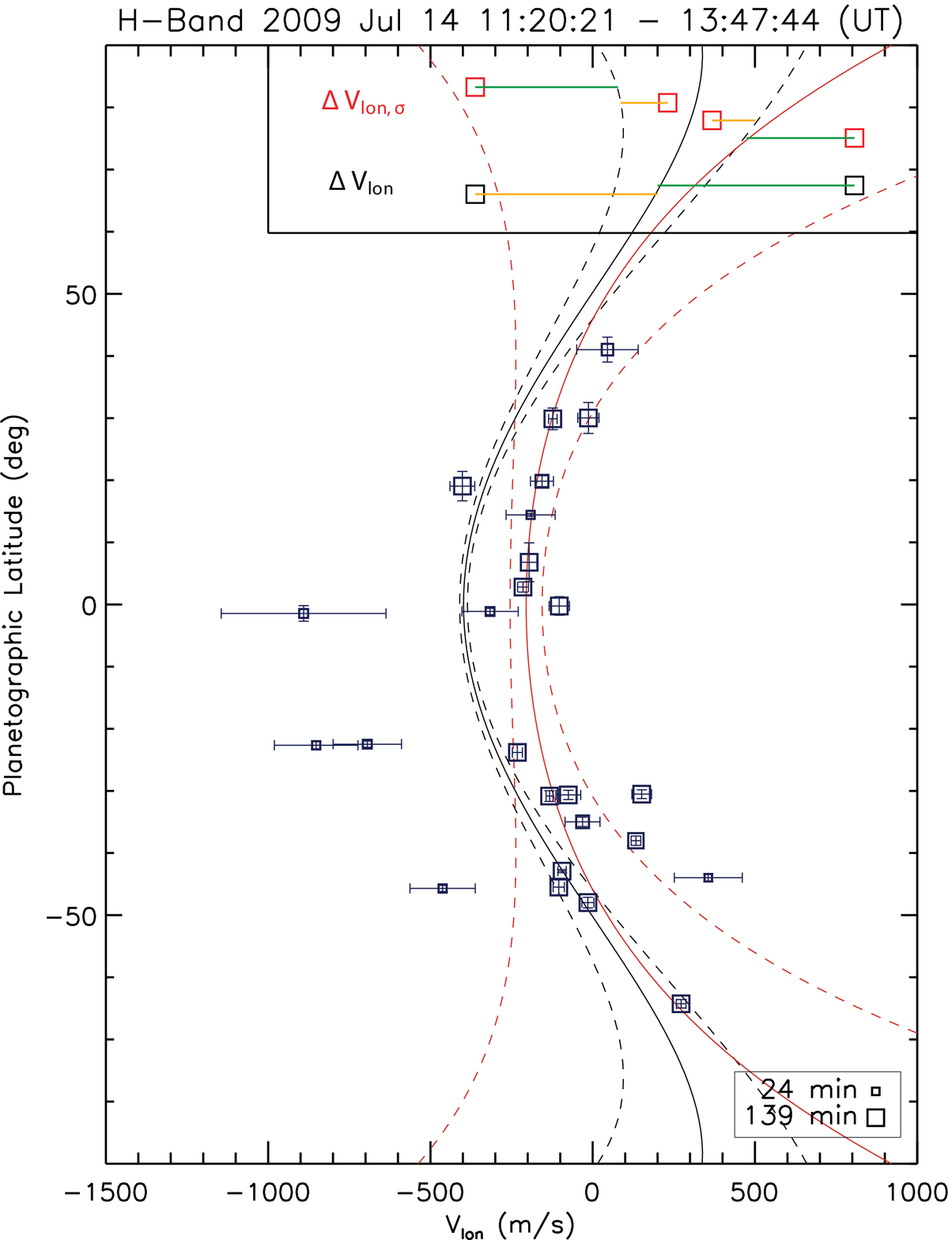}
\caption{Derived zonal velocities of 14 July H-band Low-$\sigma_{rl}$ features shown against the smooth Voyager wind
    profile (\textbf{black solid line}).  The
width of uncertainty in the smooth Voyager wind profile is shown with \textbf{black dashed lines}.  The length of time over which each
feature was tracked is indicated by the \textbf{size of the square} in
which each feature's zonal velocity is plotted.  Square sizes
corresponding to the shortest and longest tracking times are shown in
the bottom left corner.  There is significant deviation and dispersion
in the zonal velocities from the smooth Voyager wind profile, greater
than what is observed on 16 July in K'-band.  Our polynomial fit to the
zonal velocities vs. latitudes of Low-$\sigma_{rl}$ features is shown
with a \textbf{red solid line}, and the widths of uncertainty of our fit
are shown with \textbf{red dashed lines}.  Our polynomial fit is listed in Table \ref{poly} along with the smooth Voyager profile fit of Sromovsky et al. (1993),
for comparison.  Our polynomial fit to the 14 July H-band feature zonal
velocity distribution deviates significantly from the smooth Voyager
wind profile, with a shift toward more positive velocities of 180$\pm$50
m/s. A graphical illustration of our two definitions of deviation in the
    zonal velocities, $\Delta V_{lon,\sigma}$ (\textbf{red ``fake''
    features}) and $\Delta V_{lon}$ (\textbf{black ``fake'' features}),
    is shown in the top legend. \textbf{Green lines} illustrate positive
    values of deviation while \textbf{yellow lines} indicate negative
    values of deviation, according to the definition of each quantity.}
\label{srovwh_h}
\end{center}
\end{figure*}

\begin{figure*}[t]
\begin{center}
\includegraphics[scale=0.7]{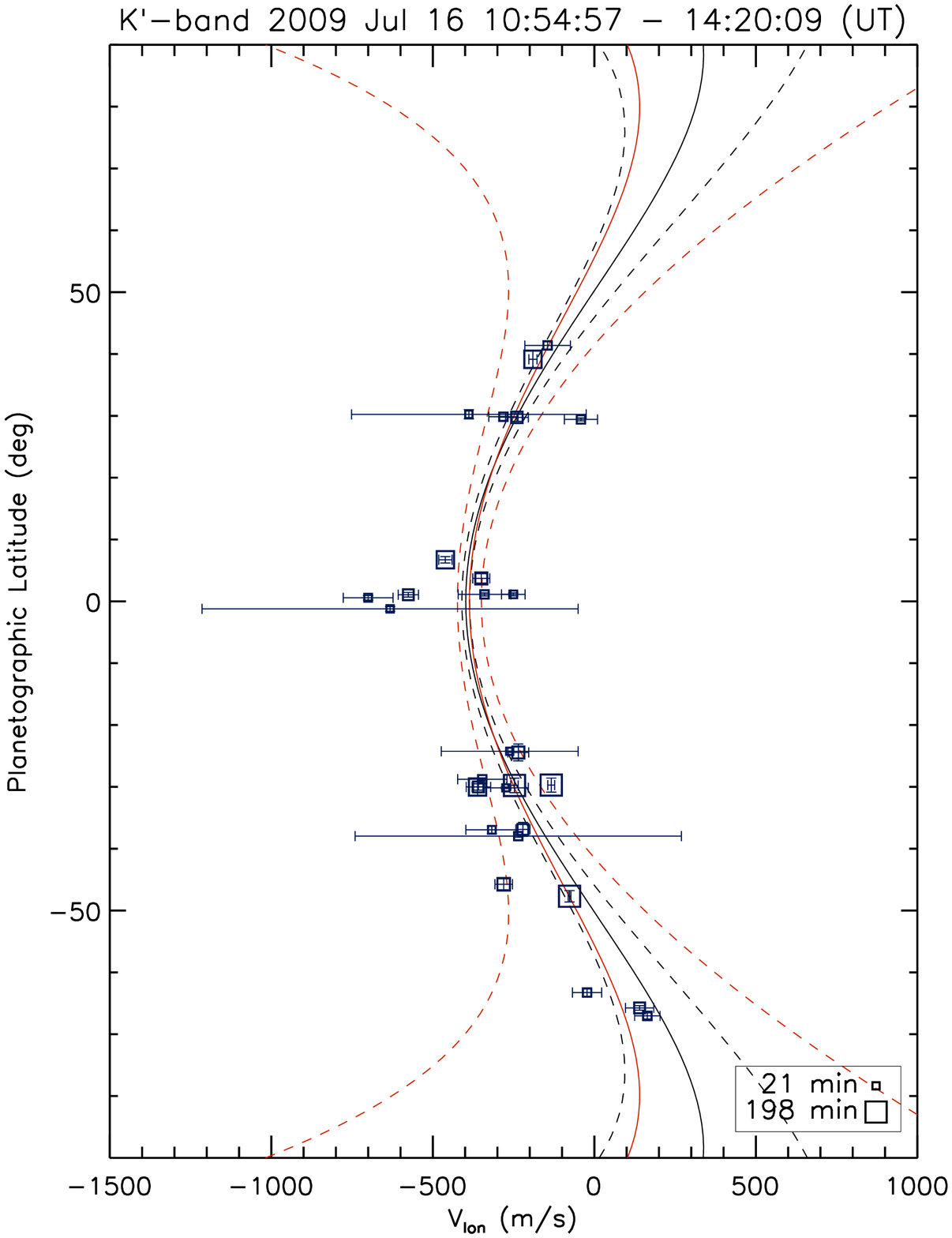}
\caption{Same as Figure \ref{srovwh_h}, now showing the derived zonal speeds of 16 July K'-band Low-$\sigma_{rl}$ features against the smooth Voyager wind
    profile (\textbf{black solid line}).  There is significant deviation
    and dispersion in zonal velocities from the smooth Voyager wind
    profile.  16 July K'-band feature zonal velocities agree well
    (within 1$\sigma$) with the smooth Voyager wind profile in their
    polynomial fit (\textbf{red solid line}).}
\label{srovwh_kp}
\end{center}
\end{figure*}

\begin{figure*}[t]
\begin{center}
\includegraphics[scale=0.7]{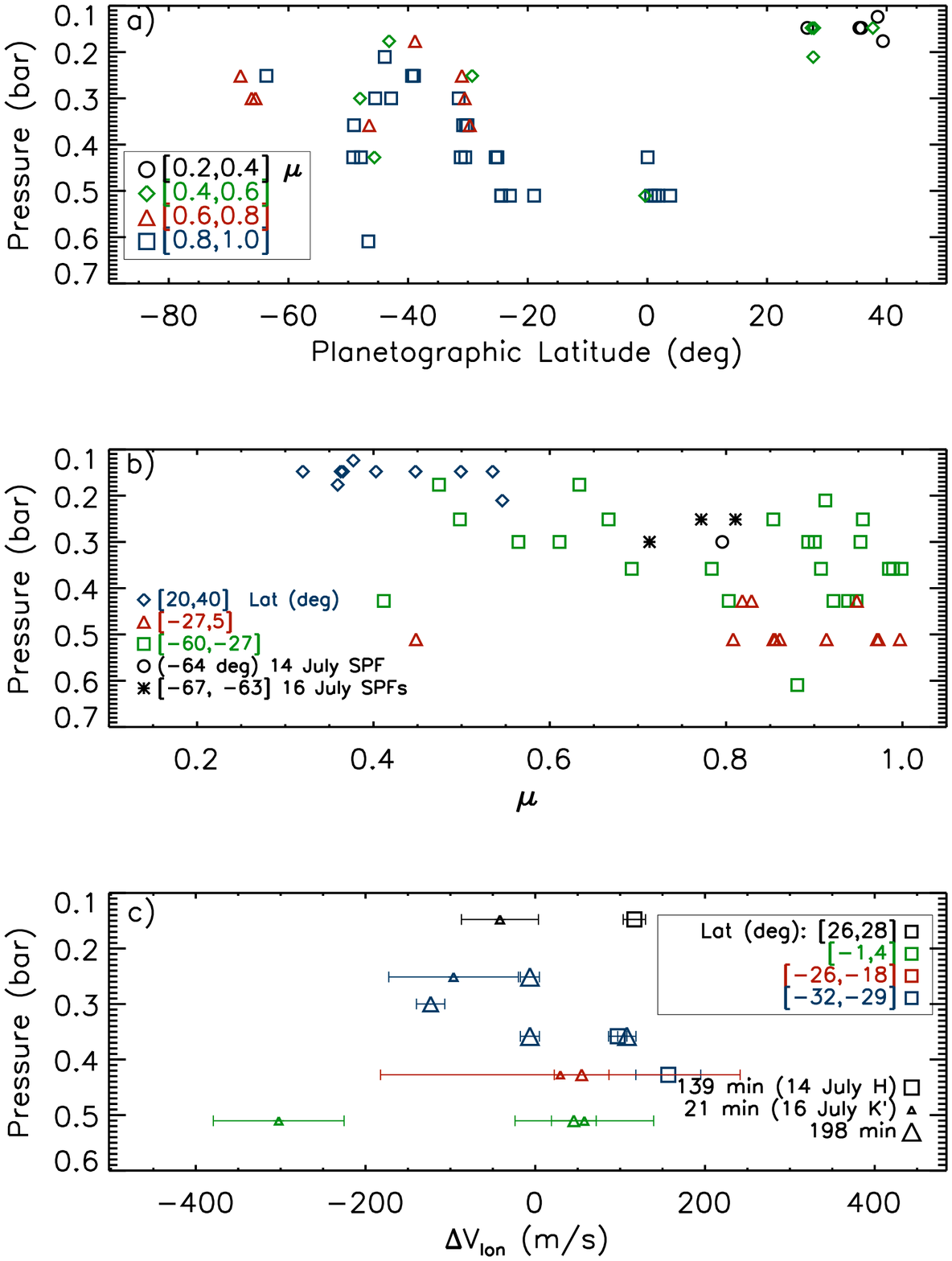}
\caption{\textbf{a)} cloud-top pressures versus latitudes
of features seen in both H- and K'-bands on 14 and 16 July.  Features
are shown in separate bins of the cosine of emission angle, $\mu$, by the
\textbf{color} and \textbf{symbol} in which they are plotted, as indicated in the bottom
left corner of the panel.  \textbf{b)} cloud-top pressures
versus $\mu$ of features shown in separate bins of latitude by the \textbf{color}
and \textbf{symbol} in which they are plotted, as indicated in the bottom left
corner of the panel.  Aside from one feature at 50$^{\circ}$S, equatorial features are uniformly deeper while features in the north are found at higher altitudes.  \textbf{c)} the cloud-top pressures vs. residuals in zonal velocity from the smooth Voyager wind profile, $\Delta V_{lon}$, of features for which we were able to make both measurements, from 32$^{\circ}$S to 28$^{\circ}$N (where vertical wind shear is expected to be important) in thin bins of latitude.  \textbf{Symbol size} indicates the length of time over which a feature was tracked, increasing linearly from the shortest to the longest tracking times indicated in the bottom right corner of the panel.  \textbf{Squares} are used to represent features whose velocities were obtained by tracking positions in 14 July H-band images and \textbf{triangles} are used to represent features whose velocities were obtained by tracking positions in 16 July K'-band images.}
\label{deep}
\end{center}
\end{figure*}

\begin{figure*}[t]
\includegraphics[width=\linewidth]{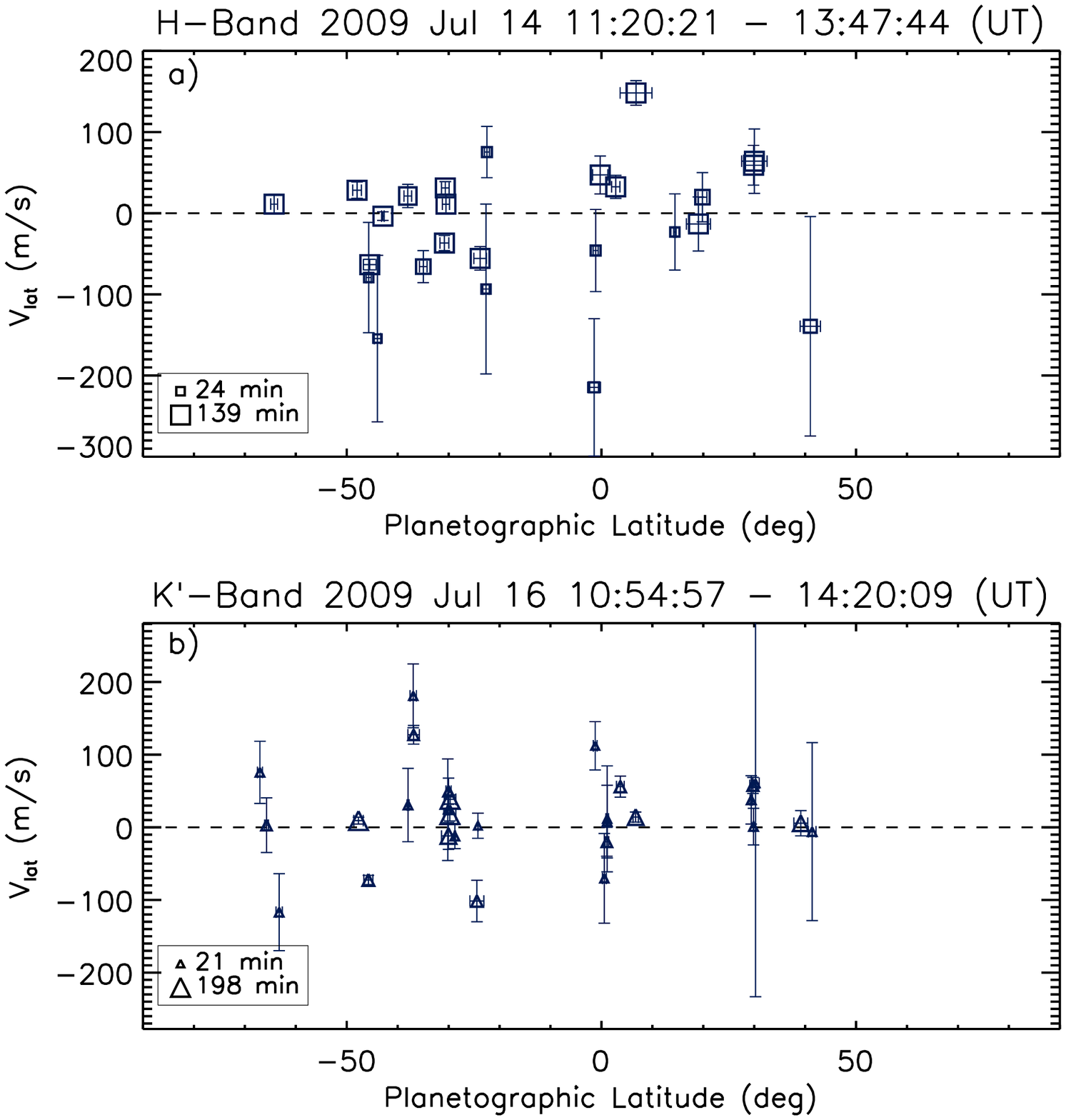}
\caption{Derived north-south wind speeds $V_{lat}$ versus mean latitude
    positions of 14 July H- (panel \textbf{a}) and 16 July K'-band
    (panel \textbf{b}) Low-$\sigma_{rl}$ features.  Length of time over which features were tracked is indicated by the
\textbf{size of the symbol} in which they are plotted, increasing
linearly from the shortest to longest tracking times indicated in each
panel.} 
\label{latspeeds}
\end{figure*}


\begin{figure*}[t]
\begin{center}
\includegraphics[scale=0.7]{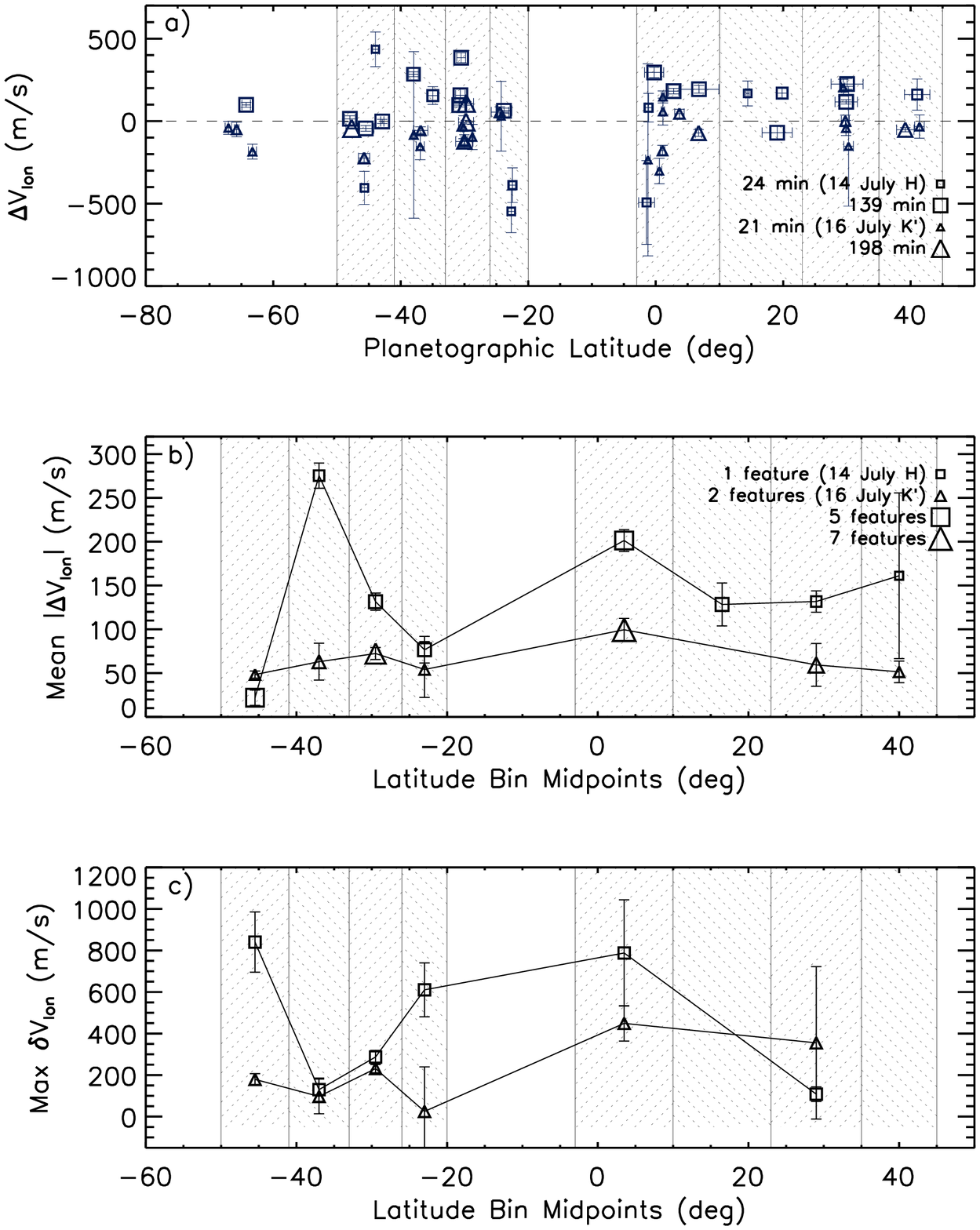}
\caption{\textbf{a)} residuals in zonal velocity from the
    smooth Voyager wind profile, $\Delta V_{lon}$, for H- (\textbf{squares}) and K'-band
    (\textbf{triangles}) Low-$\sigma_{rl}$ features.  Length of time over which features were
    tracked is indicated by the \textbf{symbol size} in which each
    feature is plotted, increasing linearly from the shortest to the
    longest tracking times in each filter, as indicated in the panel.  Latitude bins in which we average
    the absolute residuals in zonal velocity of features are indicated
    by \textbf{shaded bars}. \textbf{b)} the mean absolute residuals in
    zonal velocity from the smooth Voyager wind profile, mean $\left|
    \Delta V_{lon} \right|$, averaged in the latitude bins indicated,
    for 14 July H- and 16 July K'-band Low-$\sigma_{rl}$ features.
    \textbf{Symbol size} indicates the number of features composing the
    mean average, increasing linearly with the number of features.
    H-band features on average show greater deviation from the smooth
    Voyager wind profile. \textbf{c)} Differences between the maximum
    and minimum zonal velocities of Low-$\sigma_{rl}$ features within
    each latitude bin (where two or more features are found in each
    filter), max $\delta V_{lon}$.}
\label{pans}
\end{center}
\end{figure*}


\begin{sidewaystable*}[h]
\begin{center}
\begin{tabular}{lllllllll}
\hline \hline
Latitude (deg) & $d \phi/dt$ (deg/hr) & $V_{lon}$ (m/s) & $\Delta V_{lon,\sigma}$ (m/s) & $d \theta/dt$ (deg/hr) & $V_{lat}$ (m/s) & $\sigma_{rl}$ (deg) & $\Delta t$ (min) & N$_{meas}$ \\ \hline
-64.3$\pm$.7 & 5.22$\pm$0.28 & 272$\pm$15 & -22$\pm$15 & 0.09$\pm$0.10 & 11$\pm$12 & 0.46 & 139 & 15 \\
-48.0$\pm$.9 & -0.18$\pm$0.23 & -15$\pm$19 & -46$\pm$19 & 0.24$\pm$0.09 & 28$\pm$10 & 0.39 & 139 & 15 \\
-45.7$\pm$.5 & -5.51$\pm$1.20 & -462$\pm$101 & 350$\pm$101 & -0.67$\pm$0.57 & -79$\pm$68 & 0.21 & 33 & 5 \\
-45.5$\pm$.3 & -1.24$\pm$0.20 & -105$\pm$17 & -10$\pm$17 & -0.53$\pm$0.05 & -63$\pm$6 & 0.25 & 139 & 15 \\
-44.0$\pm$.6 & 4.12$\pm$1.21 & 356$\pm$105 & 385$\pm$105 & -1.30$\pm$0.86 & -154$\pm$103 & 0.16 & 24 & 4 \\
-42.9$\pm$.3 & -1.07$\pm$0.14 & -94$\pm$13 & -46$\pm$13 & -0.03$\pm$0.04 & -4$\pm$5 & 0.27 & 139 & 15 \\
-38.0$\pm$.8 & 1.41$\pm$0.16 & 133$\pm$15 & 246$\pm$15 & 0.18$\pm$0.12 & 21$\pm$14 & 0.25 & 123 & 13 \\
-35.0$\pm$.8 & -0.32$\pm$0.55 & -31$\pm$54 & 121$\pm$54 & -0.55$\pm$0.17 & -66$\pm$20 & 0.38 & 91 & 9 \\
-30.8$\pm$.9 & -1.29$\pm$0.10 & -133$\pm$11 & 69$\pm$11 & -0.31$\pm$0.08 & -37$\pm$9 & 0.17 & 139 & 15 \\
-30.6$\pm$.7 & -0.73$\pm$0.37 & -75$\pm$38 & 129$\pm$38 & 0.26$\pm$0.07 & 31$\pm$8 & 0.61 & 139 & 15 \\
-30.5$\pm$.7 & 1.46$\pm$0.29 & 151$\pm$30 & 357$\pm$30 & 0.09$\pm$0.10 & 11$\pm$12 & 0.36 & 139 & 15 \\
-23.8$\pm$.2 & -2.11$\pm$0.14 & -232$\pm$15 & 42$\pm$15 & -0.47$\pm$0.12 & -56$\pm$14 & 0.19 & 139 & 15 \\
-22.7$\pm$.6 & -7.69$\pm$1.16 & -852$\pm$129 & 527$\pm$129 & -0.79$\pm$0.88 & -93$\pm$105 & 0.20 & 33 & 5 \\
-22.5$\pm$.5 & -6.26$\pm$0.95 & -694$\pm$105 & 369$\pm$105 & 0.64$\pm$0.27 & 75$\pm$32 & 0.24 & 41 & 6 \\
-1.4$\pm$.3 & -7.42$\pm$2.12 & -891$\pm$254 & 481$\pm$254 & -1.82$\pm$0.72 & -215$\pm$85 & 0.53 & 41 & 6 \\
-1.1$\pm$.5 & -2.63$\pm$0.72 & -316$\pm$87 & 69$\pm$87 & -0.39$\pm$0.43 & -46$\pm$51 & 0.19 & 41 & 6 \\
-0.2$\pm$.5 & -0.86$\pm$0.26 & -103$\pm$31 & 283$\pm$31 & 0.40$\pm$0.20 & 47$\pm$23 & 0.39 & 139 & 15 \\
2.8$\pm$.8 & -1.79$\pm$0.14 & -215$\pm$17 & 170$\pm$17 & 0.28$\pm$0.12 & 33$\pm$14 & 0.23 & 139 & 15 \\
6.8$\pm$.1 & -1.64$\pm$0.20 & -196$\pm$24 & 181$\pm$24 & 1.26$\pm$0.13 & 148$\pm$15 & 0.35 & 139 & 15 \\
14.4$\pm$.3 & -1.65$\pm$0.65 & -191$\pm$76 & 153$\pm$76 & -0.20$\pm$0.40 & -23$\pm$47 & 0.09 & 33 & 5 \\
19.1$\pm$.4 & -3.54$\pm$0.34 & -401$\pm$38 & 53$\pm$38 & -0.11$\pm$0.28 & -13$\pm$33 & 0.50 & 139 & 15 \\
19.8$\pm$.9 & -1.38$\pm$0.31 & -156$\pm$35 & 152$\pm$35 & 0.17$\pm$0.26 & 20$\pm$30 & 0.25 & 91 & 9 \\
29.9$\pm$.8 & -1.18$\pm$0.13 & -123$\pm$13 & 90$\pm$13 & 0.50$\pm$0.21 & 59$\pm$24 & 0.19 & 139 & 15 \\
30.0$\pm$.5 & -0.13$\pm$0.32 & -13$\pm$33 & 198$\pm$33 & 0.54$\pm$0.33 & 64$\pm$40 & 0.42 & 139 & 15 \\
41.0$\pm$.1 & 0.50$\pm$1.04 & 45$\pm$95 & 117$\pm$95 & -1.17$\pm$1.14 & -139$\pm$135 & 0.38 & 74 & 7 \\
\hline
\end{tabular}
\end{center}
\caption{Table of various values for Low-$\sigma_{rl}$ features tracked in H-band on 14
  July. Definitions of most quantities are given throughout the text. $\Delta t$ is the length of time over which each feature was
  tracked, and $N_{meas}$ is the number of images in which each feature
  was measured. For a description of the quantity $\Delta V_{lon,\sigma}$ see footnote 2.}
\label{htab}
\end{sidewaystable*}

\begin{sidewaystable*}[h]
\begin{center}
\begin{tabular}{lllllllll}
\hline \hline
Latitude (deg) & $d \phi/dt$ (deg/hr) & $V_{lon}$ (m/s) & $\Delta V_{lon,\sigma}$ (m/s) & $d \theta/dt$ (deg/hr) & $V_{lat}$ (m/s) & $\sigma_{rl}$ (deg) & $\Delta t$ (min) & N$_{meas}$ \\ \hline
-67.0$\pm$.5 & 3.50$\pm$0.84 & 164$\pm$39 & -95$\pm$39 & 0.63$\pm$0.36 & 76$\pm$43 & 0.17 & 36 & 6 \\
-65.7$\pm$.4 & 2.84$\pm$0.89 & 140$\pm$44 & -78$\pm$44 & 0.03$\pm$0.31 & 3$\pm$38 & 0.34 & 67 & 7 \\
-63.3$\pm$.6 & -0.42$\pm$0.84 & -22$\pm$45 & 69$\pm$45 & -0.98$\pm$0.44 & -117$\pm$53 & 0.16 & 36 & 6 \\
-47.7$\pm$.9 & -0.95$\pm$0.05 & -77$\pm$4 & -15$\pm$4 & 0.08$\pm$0.04 & 9$\pm$5 & 0.15 & 198 & 21 \\
-45.8$\pm$.9 & -3.35$\pm$0.32 & -281$\pm$27 & 169$\pm$27 & -0.61$\pm$0.05 & -72$\pm$6 & 0.21 & 89 & 10 \\
-38.0$\pm$.2 & -2.49$\pm$5.34 & -235$\pm$505 & 45$\pm$505 & 0.26$\pm$0.42 & 31$\pm$50 & 0.14 & 36 & 6 \\
-37.0$\pm$.6 & -3.31$\pm$0.84 & -317$\pm$81 & 117$\pm$81 & 1.52$\pm$0.37 & 181$\pm$44 & 0.15 & 28 & 5 \\
-36.9$\pm$.1 & -2.30$\pm$0.23 & -221$\pm$22 & 20$\pm$22 & 1.07$\pm$0.11 & 127$\pm$13 & 0.09 & 67 & 7 \\
-30.2$\pm$.1 & -2.64$\pm$0.67 & -274$\pm$70 & 10$\pm$70 & 0.20$\pm$0.59 & 24$\pm$70 & 0.05 & 21 & 4 \\
-30.0$\pm$.4 & -3.48$\pm$0.16 & -361$\pm$17 & 96$\pm$17 & -0.1$\pm$0.15 & -12$\pm$18 & 0.32 & 149 & 18 \\
-30.0$\pm$.5 & -3.45$\pm$0.37 & -359$\pm$38 & 93$\pm$38 & 0.42$\pm$0.15 & 50$\pm$18 & 0.16 & 67 & 7 \\
-29.8$\pm$.2 & -2.37$\pm$0.11 & -247$\pm$11 & -20$\pm$11 & 0.15$\pm$0.07 & 17$\pm$9 & 0.37 & 198 & 21 \\
-29.7$\pm$.1 & -1.28$\pm$0.10 & -133$\pm$11 & 81$\pm$11 & 0.32$\pm$0.05 & 39$\pm$6 & 0.25 & 198 & 21 \\
-28.8$\pm$.1 & -3.30$\pm$0.73 & -347$\pm$76 & 71$\pm$76 & -0.1$\pm$0.14 & -12$\pm$17 & 0.15 & 36 & 6 \\
-24.4$\pm$.4 & -2.15$\pm$0.29 & -235$\pm$32 & 33$\pm$32 & -0.86$\pm$0.24 & -101$\pm$29 & 0.20 & 82 & 9 \\
-24.2$\pm$.1 & -2.39$\pm$1.94 & -262$\pm$212 & 8$\pm$212 & 0.02$\pm$0.15 & 2$\pm$17 & 0.16 & 21 & 4 \\
-1.2$\pm$.4 & -5.27$\pm$4.85 & -632$\pm$582 & 223$\pm$582 & 0.95$\pm$0.28 & 112$\pm$33 & 0.47 & 23 & 4 \\
0.6$\pm$.4 & -5.83$\pm$0.64 & -700$\pm$77 & 290$\pm$77 & -0.60$\pm$0.52 & -70$\pm$62 & 0.12 & 28 & 5 \\
1.1$\pm$.4 & -4.80$\pm$0.26 & -576$\pm$31 & 166$\pm$31 & -0.17$\pm$0.17 & -20$\pm$20 & 0.12 & 67 & 7 \\
1.1$\pm$.3 & -2.83$\pm$0.68 & -340$\pm$82 & 46$\pm$82 & 0.07$\pm$0.42 & 8$\pm$50 & 0.13 & 36 & 6 \\
1.1$\pm$.2 & -2.09$\pm$0.30 & -251$\pm$37 & 135$\pm$37 & 0.10$\pm$0.62 & 12$\pm$73 & 0.05 & 28 & 5 \\
3.7$\pm$.8 & -2.92$\pm$0.22 & -350$\pm$26 & 33$\pm$26 & 0.48$\pm$0.12 & 56$\pm$14 & 0.14 & 82 & 9 \\
6.7$\pm$.5 & -3.87$\pm$0.17 & -461$\pm$21 & 59$\pm$21 & 0.12$\pm$0.06 & 14$\pm$7 & 0.34 & 149 & 18 \\
29.4$\pm$.2 & -0.40$\pm$0.49 & -41$\pm$51 & 177$\pm$51 & 0.32$\pm$0.28 & 38$\pm$33 & 0.11 & 36 & 6 \\
29.7$\pm$.6 & -2.30$\pm$0.34 & -240$\pm$35 & -25$\pm$35 & 0.49$\pm$0.09 & 58$\pm$11 & 0.15 & 74 & 8 \\
29.9$\pm$.1 & -2.70$\pm$0.44 & -282$\pm$45 & 15$\pm$45 & 0.01$\pm$0.21 & 1$\pm$25 & 0.09 & 36 & 6 \\
30.2$\pm$.8 & -3.74$\pm$3.50 & -388$\pm$363 & 125$\pm$363 & 0.51$\pm$2.48 & 61$\pm$294 & 0.33 & 23 & 4 \\
39.1$\pm$.40 & -2.04$\pm$0.13 & -190$\pm$13 & 12$\pm$13 & 0.05$\pm$0.15 & 6$\pm$17 & 0.33 & 149 & 18 \\
41.4$\pm$.7 & -1.61$\pm$0.78 & -145$\pm$70 & -11$\pm$70 & -0.05$\pm$1.03 & -6$\pm$123 & 0.14 & 36 & 6 \\
\hline
\end{tabular}
\end{center}
\caption{Table of various values for Low-$\sigma_{rl}$ features tracked in K'-band on 16 July. Definitions of most quantities are given throughout the text. $\Delta t$ is the length of time over which each feature was
  tracked, and $N_{meas}$ is the number of images in which each feature
  was measured. For a description of the quantity $\Delta V_{lon,\sigma}$ see footnote 2.}
\label{kptab}
\end{sidewaystable*}

\clearpage

\makeatletter
\let\clear@thebibliography@page=\relax
\makeatother

\end{document}